\documentclass[letterpaper]{JHEP3}

\usepackage{epsfig}
\usepackage{graphicx}
\usepackage{amssymb,amsmath}
\usepackage{epstopdf}

\def\be{\begin{eqnarray}}
\def\ee{\end{eqnarray}}

\title{Holographic Nuclear Physics}

\author{Oren Bergman\\
Department of Physics\\
Technion, Haifa 32000, Israel\\
\email{bergman@physics.technion.ac.il}}

\author{Gilad Lifschytz\\
Department of Mathematics and Physics and CCMSC \\
University of Haifa at Oranim\\
Tivon 36006, Israel \\
\email{giladl@research.haifa.ac.il}}

\author{Matthew Lippert\\
Department of Physics\\
Technion, Haifa 32000, Israel\\
{\rm and}\\
Department of Mathematics and Physics \\
University of Haifa at Oranim\\
Tivon 36006, Israel\\
\email{matthewslippert@gmail.com}}

\date{}

\abstract{We analyze the phases of the Sakai-Sugimoto model at finite 
temperature and baryon chemical potential.
Baryonic matter is represented either by 4-branes in the 8-branes or
by strings stretched from the 8-branes to the horizon.
We find the explicit configurations and use them to determine the phase diagram
and equation of state of the model.
The 4-brane configuration (nuclear matter) is always preferred to the string configuration
(quark matter), and the latter is also unstable to density fluctuations.
In the deconfined phase
the phase diagram has three regions corresponding to the vacuum, quark-gluon plasma,
and nuclear matter, with a first-order and a second-order phase transition
separating the phases.
We find that for a large baryon number density, 
and at low temperatures, the dominant phase has 
broken chiral symmetry.
This is in qualitative agreement with studies of QCD at 
high density.}

\begin{document}


\section{Introduction}

QCD at finite baryon density has a rich phase 
structure (for reviews see \cite{Schafer:2005ff,Alford,Rajagopal}).
Naively one would expect that at high density, like at high temperature, QCD is in
a deconfined chiral-symmetric quark-gluon plasma phase. It turns out, however, that new phases 
appear at high density, in which both the chiral symmetry and the gauge 
symmetry are broken \cite{Alford,Rajagopal}. In real QCD with $N_c=3$
and three light flavors of quarks
the dominant phase at high density is a color-flavor-locking (CFL) phase
\cite{Alford:1998mk}.
With two light flavors of quarks the dominant phase is a color-superconductor.
At large $N_c$ it is believed that these gauge-symmetry breaking phases
are suppressed, and the dominant phase at high density
is a chiral density wave \cite{dgr,ss}, in which the chiral symmetry (only) is broken
non-uniformly.
It appears therefore that QCD (both with $N_c=3$ and at large $N_c$)
at low temperature and high density always has broken chiral symmetry.

These results rely on perturbative calculations in QCD,
and analogous models such as the Nambu-Jona-Lasinio (NJL) model, at finite density
near the Fermi surface, and are therefore limited to values of the chemical 
potential for which $\alpha_s(\mu) \ll 1$.
At present, lattice QCD techniques are unable to deal with a (large) baryon chemical
potential (for a review see \cite{Lombardo:2006yc}).

At large $N_c$, gauge/gravity duality is an alternative approach to gauge theory at strong coupling
\cite{Aharony:1999ti}.
Several recent models have incorporated flavors using probe branes in backgrounds dual
to large $N_c$ Yang-Mills theories with various amounts of supersymmetry
\cite{Flavors}. 
The Sakai-Sugimoto model in particular is quite similar to QCD at large $N_c$
\cite{Sakai:2004cn}. This model builds on Witten's model for pure Yang-Mills theory in four dimensions,
which uses 4-branes wrapped on a Scherk-Schwarz circle \cite{Witten:1998zw}, and adds $N_f$
probe 8-branes and $N_f$ probe anti-8-branes transverse to the circle. 
These provide massless chiral fermions
(left-handed from the 8-branes, right-handed from the anti-8-branes)
in the fundamental representation
of both the gauge group $U(N_c)$, and the flavor group
$U(N_f)_L\times U(N_f)_R$.

One of the most compelling features of this model is that
it describes spontaneous chiral-symmetry breaking in a simple
geometrical way. Since in the near-horizon limit the circle vanishes at
a finite radial coordinate, the 8-branes and anti-8-branes are smoothly
connected into a U-shaped configuration with an asymptotic separation $L$
at infinity. The actual embedding of the 8-branes is determined by solving the DBI
equations of motion with this boundary condition.

The model also exhibits many other properties similar to QCD 
\cite{SS_model_general,Hata:2007mb}.
In particular it has an interesting phase structure at finite temperature
\cite{Aharony:2006da}. At low temperature the model is essentially the same as at zero
temperature, {\em i.e.} it describes a confining gauge theory with broken chiral symmetry.
At high temperature the model deconfines and chiral symmetry is restored,
which is described geometrically by the separation of the 8-branes and anti-8-branes.
For sufficiently small $L$ there is also an intermediate range of temperatures at which the model
is deconfined but chiral symmetry remains broken.
In the deconfined phase both the connected U-configuration and the separated
parallel configuration are possible. The dominant configuration, and therefore phase,
is determined by comparing their actions.

The baryonic $U(1)_V$ symmetry corresponds in models with fundamental matter
to the diagonal $U(1)$ gauge symmetry of the probe branes.
Baryon number is therefore described by electric charge, 
and baryon number density is related to the electric field, or
more precisely to the electric displacement field, of the diagonal $U(1)$. 
Correspondingly, the baryon chemical potential is described by the value
of the gauge potential at infinity $A_0(\infty)$.
Finite baryon density in the Sakai-Sugimoto model has been studied
in \cite{Kim:2006gp,Horigome:2006xu,Sin:2007ze,Yamada:2007ys}.
However, only part of the parameter space has been explored so far.
Other models with finite baryon density have been studied in
\cite{Kobayashi:2006sb,Domokos:2007kt,Kim:2007em,Kim:2007xi}.

\medskip

In this paper we explore the full parameter space
of the Sakai-Sugimoto model at finite temperature and
finite uniform baryon number density,
in both the confined and deconfined phases.
Other than temperature and baryon number density (or chemical potential),
this model has an additional parameter not present in QCD, namely
the asymptotic 8-brane-anti-8-brane separation $L$.
We will assume that the value of $L$ is such that the intermediate phase
of deconfinement with chiral symmetry breaking exists, in other words that
the deconfinement temperature is (much) smaller than the chiral-symmetry
restoration temperature.
The confined phase is, of course, of great interest, but the deconfined phase
exhibits a much richer phase diagram. The deconfined phase is also 
qualitatively the same as the non-local NJL model \cite{NJL,PS}, and we
expect a similar phase diagram in that case.

There are two types of objects which carry baryon charge. The baryons themselves
correspond to 4-branes wrapped on the $S^4$ part of the background. 
Due to the RR flux each 4-brane comes with $N_c$ strings attached \cite{Witten:1998xy}.
The other end of the strings is attached to the 8-branes, which is how the baryons 
get their flavor.
However, in the nonsupersymmetric 4-brane background this configuration is not static.
The strings pull the wrapped 4-branes up toward the 8-branes \cite{Callan:1999zf}.
When they reach the 8-branes, the 4-branes can also be described
as instantons in the world-volume theory of the 8-branes. 
In the deconfined phase baryon charge can also be carried by strings
which stretch from the 8-branes all the way to the horizon.\footnote{We can think
of these strings as ending on massless, wrapped 4-branes at the horizon.}
This describes a possible phase in which baryon charge is carried by
free quarks.

In both cases we will consider a uniform distribution in $\mathbb{R}^3$ of baryon charge,
so the 8-brane worldvolume theory reduces to a one-dimensional problem in the
radial coordinate with a source term.  As the 4-branes and strings exert a force on the 8-branes, 
their embedding at finite baryon number density will have a cusp.

We will make the following approximations. 
First, we will assume that the wrapped 4-branes are pointlike in the 
transverse coordinates and uniformly distributed in $\mathbb{R}^3$.
The precise description would be in terms of instantons in the 8-brane
worldvolume theory.  However, instanton solutions in DBI are not known. 
Analysis in the Yang-Mills (plus Chern-Simons) approximation shows that an instanton 
has a finite size on the order of the string length \cite{Hata:2007mb}.
Using pointlike instantons, or equivalently pointlike 4-branes, is therefore
a good approximation.
We will also neglect any direct 
interactions between the 4-branes themselves or between the strings.

\begin{figure}
\centerline{\epsfig{file= 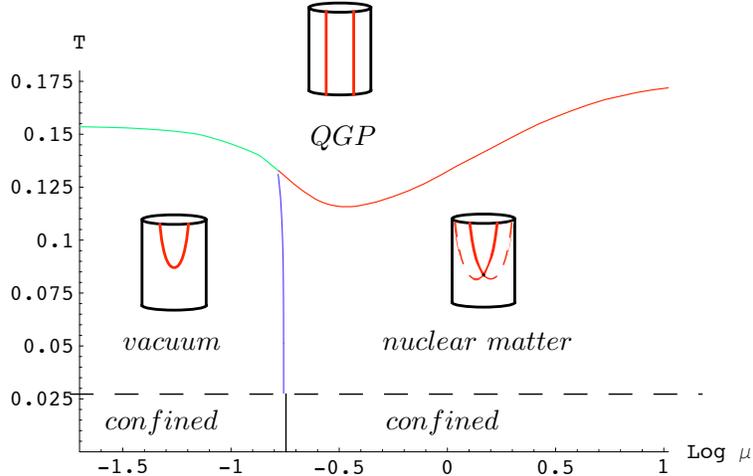,width=10cm}}
\vspace{-5cm}
\caption{The phases of holographic QCD at finite temperature and baryon 
chemical potential. A particular deconfinement temperature (0.025) was chosen for illustration
purpose only.}
\label{full_phase_diagram}
\end{figure}

Our results are summarized in the phase diagram in figure \ref{full_phase_diagram}. 
At low temperatures the theory confines, and there is a second order
phase transition at finite $\mu$ to a phase of nuclear matter.
At high temperature the theory deconfines and chiral symmetry is restored.
At intermediate temperatures chiral symmetry is broken for all $\mu$ (as in QCD).
We find a second order phase transition also in the intermediate temperature
range between the vacuum and nuclear matter phases.
This is similar to QCD, but in QCD it is a first-order transition due to 
the attractive interaction between the baryons (which we have neglected).

In section 2 we begin by describing the possible 8-brane configurations corresponding
to the Sakai-Sugimoto model at finite baryon number density.
In section 3 we discuss the thermodynamics of the gauge theory
which are implied by these configurations and derive the full phase diagram
in the grand canonical ensemble.
We consider both the confined and deconfined phases, and both
4-branes and strings as sources of baryon charge in the deconfined phase.
It will turn out that 4-branes are always preferred to strings, and that the stringy 
``quark matter'' phase is actually unstable to density fluctuations.
We conclude and offer suggestions for future work in section 4.



\section{Finite density brane configurations}

The basic brane configuration consists of $N_f$ 8-branes and $N_f$ anti-8-branes 
in the near horizon background of $N_c$ 4-branes wrapped on a Scherk-Schwarz
circle with $N_c\gg N_f$.
At zero temperature the background is capped and the circle is topologically trivial,
so the 8-branes and anti-8-branes connect into a U-shaped configuration.
The dual gauge theory is confining, and chiral symmetry is broken.
At finite temperature this continues to be the only possible configuration until
one reaches a critical temperature, at which the dominant 
background switches to the black hole, and both U-shaped 8-branes and separated parallel
8-branes and anti-8-branes are allowed. 
At first the U-shaped configuration dominates, so chiral symmetry remains broken even though
the gauge theory deconfines. 
Chiral symmetry restoration occurs at a second critical temperature,
which for $L$ small enough, is above 
the first critical temperature (otherwise they are equal).
At this temperature the separated 8-brane-anti-8-brane configuration begins to dominate.

The baryon number current is related holographically to the diagonal $U(1)$
part of the 8-brane gauge field. To study finite baryon number density configurations we
therefore need to include this gauge field in the 8-brane action.
The first place it enters is in the DBI action: 
\be
S_{D8} = - \mu_8 \int d^9 X \, e^{-\phi}\, \mbox{Tr} \,
\sqrt{-\mbox{det} (g_{MN} +  2\pi\alpha'{\cal F}_{MN})} 
\ee
where ${\cal F}$ is the $U(N_f)$ field strength
\be
{\cal F} = d{\cal A} + i{\cal A}\wedge{\cal A} \,.
\ee
We decompose the $U(N_f)$ gauge field into an $SU(N_f)$ part and a $U(1)$
part as follows\footnote{We are using the convention $\mbox{Tr}\, T_a T_b = {1\over 2}\delta_{ab}$.}
\be
{\cal A}  = A + {1\over \sqrt{2N_f}}\hat{A} \,.
\ee
The $U(1)$ gauge field will also appear in the CS action, which will be important below.
Let's study the effect of turning on this gauge field on the brane configuration in the
different phases.

\subsection{confined phase}

In the confined phase the background (at finite temperature) is given by
\be
\label{confined_background}
ds^2 &=& \left({U\over R}\right)^{3\over 2} \left((dX_0^E)^2 + (d{\bf X})^2
+ f(U)dX_4^2\right)
+ \left({R\over U}\right)^{3\over 2}\left({dU^2\over f(U)}
+ U^2 d\Omega_4^2\right) \\
e^{\Phi} &=& g_s \left({U\over R}\right)^{3/4} \\
F_4 &=& {(2\pi)^3 (\alpha')^{3/2}  N_c\over \Omega_4}\, \epsilon_4 
\ee
where $X_0^E\sim X_0^E + \beta$, $X_4\sim X_4 + \beta_4$, and
\be
f(U) = 1 - {U^3_{KK}\over U^3} \;\; , \;\;
U_{KK} = \left({4\pi\over 3}\right)^2 {R^3\over\beta_4^2} \;\; , \;\;
R^3 = \pi g_s N_c (\alpha')^{3/2} \,.
\ee
It is convenient to express everything in terms of dimensionless quanitities, so we define
\be
u={U\over R}\; ,\; x_4 = {X_4\over R} \; , \; \tau = {X_0^E\over R} \; , \; 
\hat{a} = {2\pi\alpha'\hat{A}\over \sqrt{2N_f} R} \,.
\ee
The 8-brane action with the $U(1)$ gauge field is then given by\footnote{This is
the action for just the 8-branes, {\em i.e.} for 1/2 of the full configuration.
The lower limit of the integral is the lowest radial position of the 8-brane
configuration, and the upper limit is infinity.}
\be 
\label{D8_action_confined}
S_{D8} =  {\cal N}
\int du \, u^4 
\left[ f(u) (x_4^\prime(u))^2 
 + {1\over u^3}
    \left( {1\over f(u)} - (\hat{a}_0^\prime(u))^2\right)
\right]^{1\over 2} \,,
\ee
where we have defined the overall normalization as
\be
{\cal N} \equiv {\mu_8 N_f \Omega_4 V_3 \beta R^5 \over g_s}  \,,
\ee
where $\Omega_4$ is the volume of a unit $S^4$, and $V_3$ is the volume of
space ($\mathbb{R}^3$). Note that the action scales as $N_f N_c$.

As will become clear in the next section it is convenient to also define 
the Legendre-transformed action 
\be
\label{transformed_D8_action_confined}
\tilde{S}_{D8} = S_{D8} + {\cal N}\int du\, d(u) \hat{a}_0^\prime(u) 
\ee
where $d(u)$ is the electric displacement field defined by
\be
\label{d_confined}
d(u) \equiv  \mbox{} - {1\over {\cal N}}\, {\delta S_{D8}\over \delta \hat{a}_0^\prime(u)} 
= {u \hat{a}_0^\prime(u)\over 
\left[ f(u) (x_4^\prime(u))^2 
 + u^{-3} \left( {1\over f(u)} - (\hat{a}_0^\prime(u))^2\right)
\right]^{1\over 2}} \,.
\ee
This gives
\be
\label{tilde_action}
\tilde{S}_{D8}=  {\cal N} \int du\, u^4
\left[f(u)(x_4^\prime(u))^2 + {1\over u^3 f(u)} \right]^{1\over 2} 
\left[ 1 +  {(d(u))^2\over u^5} \right]^{1\over 2} \,.
\ee
The equations of motion for $x_4(u)$ and $d(u)$ can be integrated once 
yielding two constants:
\be 
d(u) &=& d \nonumber \\[5pt]
(x_4^\prime(u))^2 &=& {1\over u^3 (f(u))^2}
\left[{f(u)(u^8+u^3d^2)\over f(u_0)(u_0^8+u_0^3d^2)} - 1 \right]^{-1} \,,
\ee
where $u_0$ is defined as the position where $x_4'(u)$ diverges.

\begin{figure}
\centerline{\epsfig{file=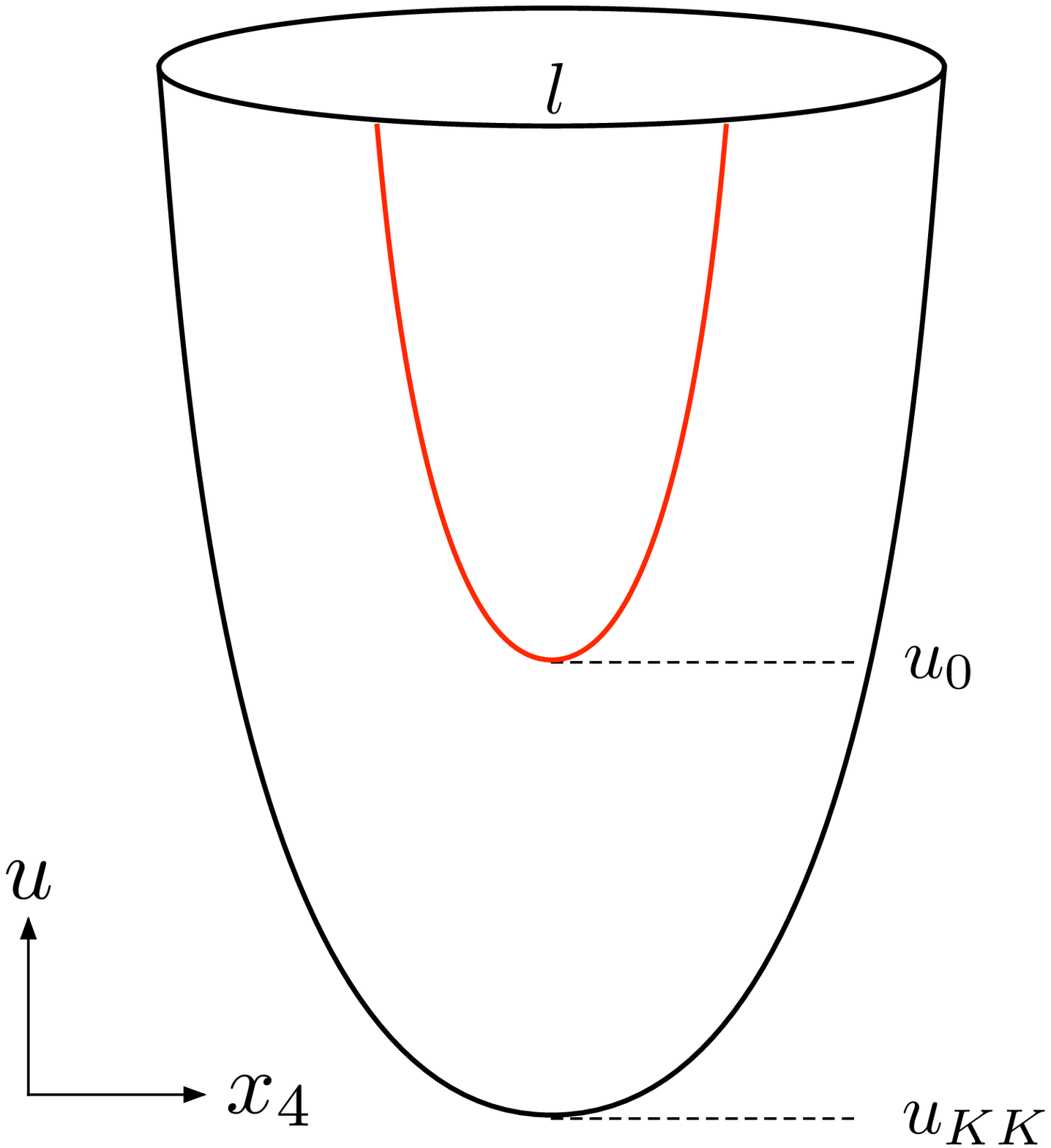,height=4cm}\hspace{1cm}\epsfig{file=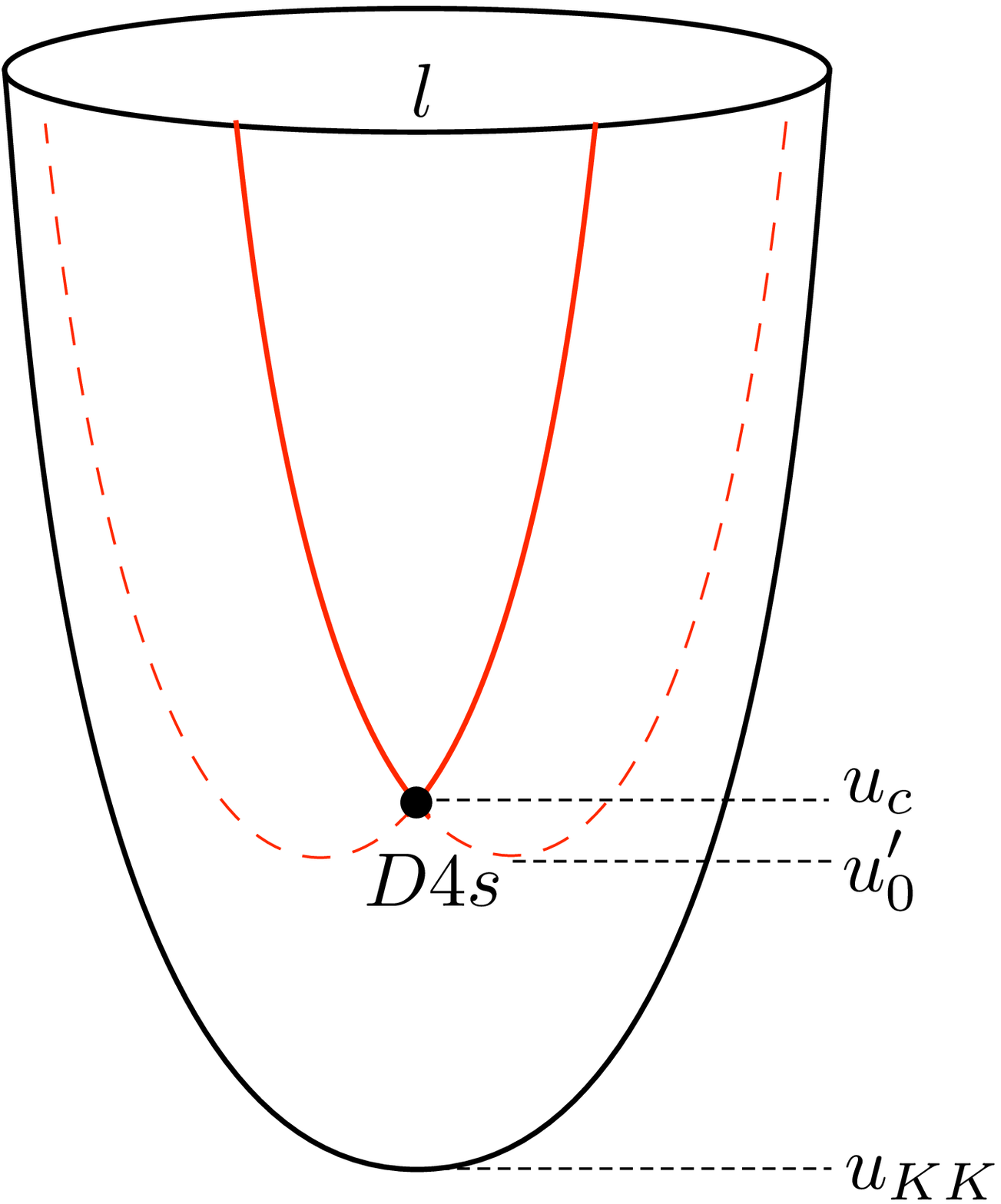,height=4cm}}
\caption{The 8-brane configuration with $d=0$ and $d\neq 0$ in the confined phase.}
\label{confined_configurations}
\end{figure}

For $d=0$ the solution is a U-shaped 8-brane in the $(x_4,u)$ plane,
with $u_0$ as its lowest radial position (figure \ref{confined_configurations})
\cite{Sakai:2004cn}.
However, a non-trivial electric displacement $d$ requires a source
at $u=u_c$, which is possibly different from $u_0$, which will
change the 8-brane configuration.
This is essentially a one-dimensional electrostatics problem in the coordinate $u$,
except that the 8-brane covers the region $[u_c,\infty]$ twice.
Each part carries an electric displacement $d$.

The only possible sources for $d$ in the confined phase 
are instantons, or equivalently 4-branes wrapped on the $S^4$,
in the 8-branes. For a uniform $d$ we need a uniform distribution in
$\mathbb{R}^3$ of 4-branes.
The source term comes from the 8-brane CS action:
\be
S_{CS} = {\mu_8\over 6}\int_{R^4\times R_+\times S^4} C_3 
\mbox{Tr}\, (2\pi\alpha' {\cal F})^3 =
{N_c\over 24\pi^2} \int_{R^4\times R_+} \omega_5({\cal A}) \,.
\ee
The relevant term is the one that couples the $U(1)$ to the $SU(N_f)$:
\be
 {N_c\over 24\pi^2} \int_{R^4\times R_+} {3\over\sqrt{2N_f}}\hat{A}_0 \mbox{Tr}\, F^2 \,.
\ee
We will assume a uniform distribution of 4-branes
in $\mathbb{R}^3$ at $u=u_c$:
\be
\label{instantons}
{1\over 8\pi^2} \mbox{Tr}\, F^2 = n_{4} \delta(u-u_c) d^3{\bf x}\, du \,,
\ee
where $n_4$ is the (dimensionless) density of 4-branes wrapped on $S^4$.
The equation of motion for the $U(1)$ gauge field then gives
\be
d'(u) = {\beta V_3 N_c\over 2\pi\alpha' R^2 {\cal N}}\, n_4\, \delta(u-u_c) \,,
\ee
and therefore\footnote{This is the density of 4-branes on one half of the 8-brane
configuration. The total 4-brane density is twice this much.}
\be 
\label{four_brane_density}
n_4 = {2\pi\alpha' R^2{\cal N}\over \beta V_3 N_c}\, d \,.
\ee

The instanton distribution (\ref{instantons}) also sources the equation of motion
for $x_4(u)$ and will therefore deform the shape of the 8-brane.
Physically, the 4-branes pull down on the 8-branes.
Since the 4-brane distribution has a finite energy density per unit 7-volume
(the $S^4$ they wrap plus the $\mathbb{R}^3$), it will form a cusp in the 8-brane
(like a bead on a string).
Away from the cusp the 8-brane will follow two opposite pieces of a U-shaped 
solution, which are truncated at some radial position $u_c$ above $u_0$
(figure \ref{confined_configurations}).

The value of $u_c$ can determined by the zero-force condition in the $(x_4,u)$ plane.
The proper tension of the 8-brane is given by varying 
the Legendre-transformed action (\ref{tilde_action}) (which is the same
as the Hamiltonian once we substitute in for the solution of $x_4(u)$)
with repsect to the proper distance along the 8-brane. The result is
\be
f_{D8} = {\cal N} u_c^{13/4} 
 \left(1 + {d^2\over u_c^5}\right)^{1/2}  \,.
\ee
The force due to the 4-branes is given simply by varying their action with respect to
their position $u_c$, again taking care to vary with respect to the proper distance.
The 4-brane action is
\be
\label{4-brane_action_confined}
S_{D4} &=&  {n_4 V_3 \mu_4\over R^3} \int d\Omega_4 d\tau \, e^{-\Phi} \, \sqrt{\mbox{det} g_{MN}} \nonumber \\
&=& {1\over 3} {\cal N} u_c d \,,
\ee
and the force is therefore
\be
f_{D4} = {\partial S_{D4}\over \partial u_c} 
\left.{1\over \sqrt{g_{uu}}}\right|_{u=u_c}
= {1\over 3} {\cal N} d u_c^{3/4} \sqrt{f(u_c)} \,.
\ee 
The condition for equilibrium is then
\be
\label{brane_equilibrium}
f_{D8}\cos\theta = f_{D4} \,,
\ee
where $\theta$ is the proper angle of the 8-brane at $u_c$,
\be
\label{cosine}
\cos\theta = \left[ 1 - {f(u_0) \left(u_0^8+u_0^3d^2\right)
 \over  f(u_c) \left(u_c^8 + u_c^3d^2 \right)}\right]^{1/2} \,.
\ee
An elegant alternative derivation of this result is given in the Appendix.

\begin{figure}
\centerline{\epsfig{file=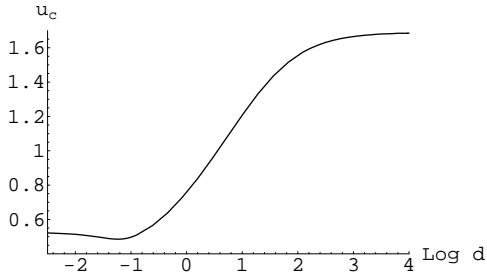,height=4cm}}
\caption{The position of the cusp (and 4-brane) in the 8-brane as a function of the 
electric displacement $d$ for $l=1$ in the confined phase. We present this as a log 
plot to show both the initial decrease,
as well as the limiting value at large $d$.}
\label{confined_uc}
\end{figure}

We want to solve this for $u_c$, while holding fixed the asymptotic separation of
the 8-branes and anti-8-branes, which is given by
\be
\label{separation}
l = 2\,\int_{u_c}^\infty du\, x_4^\prime(u) \,.
\ee
This can be done numerically by varying $u_0$ and $d$, computing $u_c$ and $l$
using (\ref{brane_equilibrium}) and (\ref{separation}), and then tabulating 
$(d,u_c)$ for a given value of $l$.
The result is presented in figure \ref{confined_uc}, where for definiteness we have set $l=1$.
Note that for small values of $d$ the cusp comes down as $d$ increases, but beyond
a certain value of $d$ it goes up. Initially the 4-branes pull the 8-branes down, but
eventually the 8-branes win this ``tug-of-war".
We will see the same behavior in the deconfined phase.
The initial downward motion of the cusp indicates that the chiral condensate
of the gauge theory decreases initially as the density increases, and
the eventual upward motion indicates that the chiral condensate eventually increases
with $d$.
As we will soon see this has an important
implication for the gauge theory thermodynamics at high density.

\subsection{deconfined phase}

\begin{figure}
\centerline{\epsfig{file=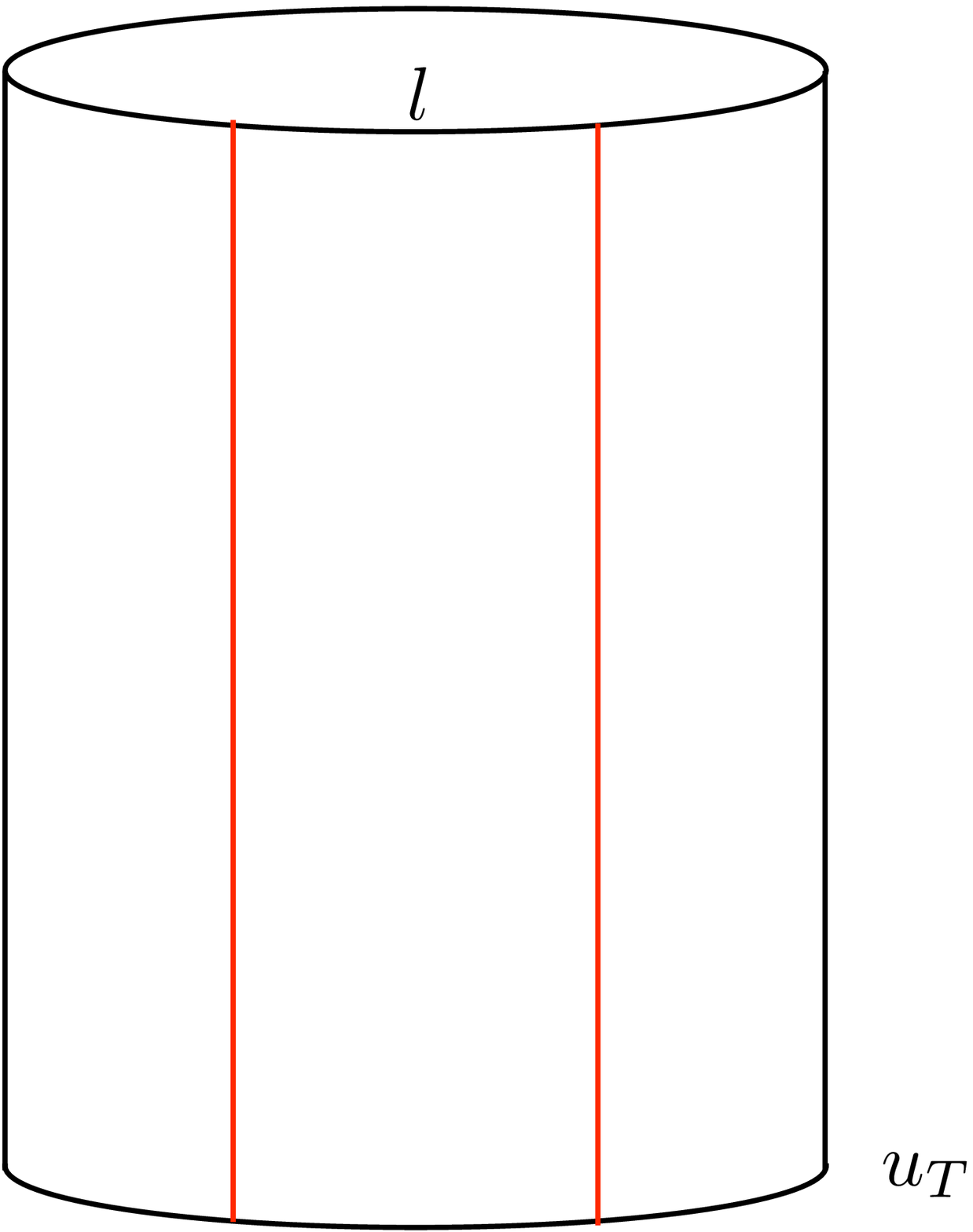,height=4cm}\hspace{0.5cm}
\epsfig{file=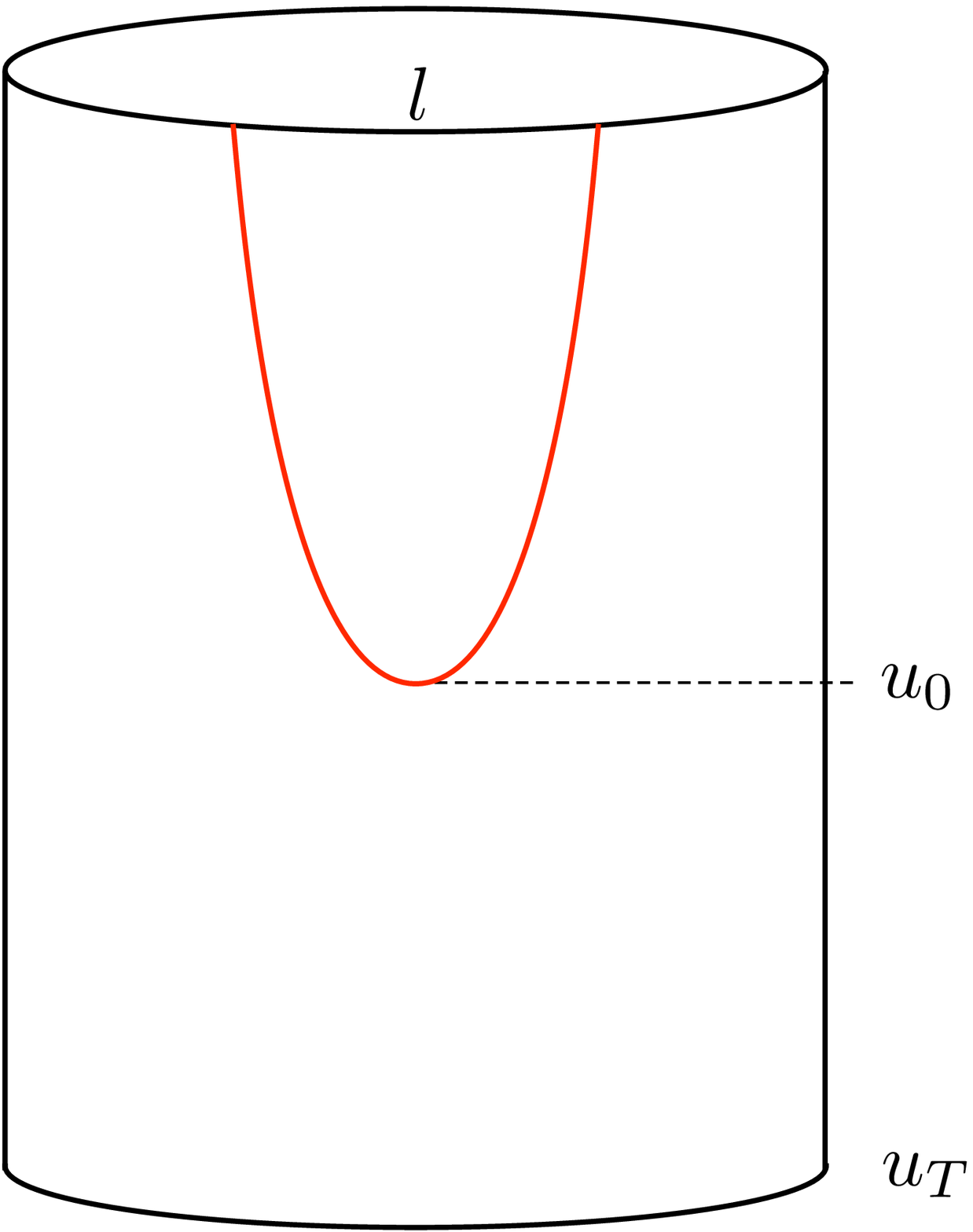,height=4cm}\hspace{0.5cm}
\epsfig{file=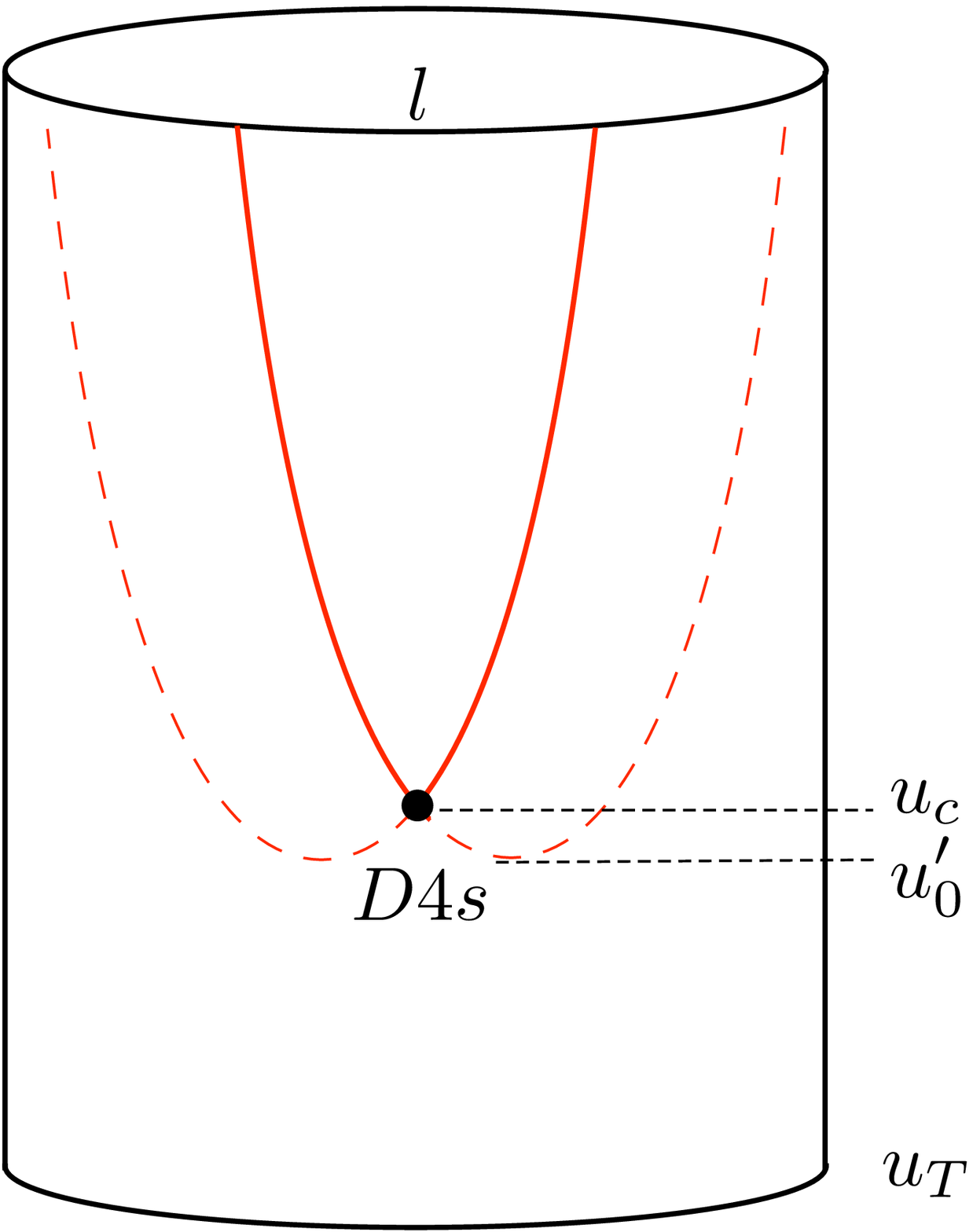,height=4cm}\hspace{0.5cm}
\epsfig{file=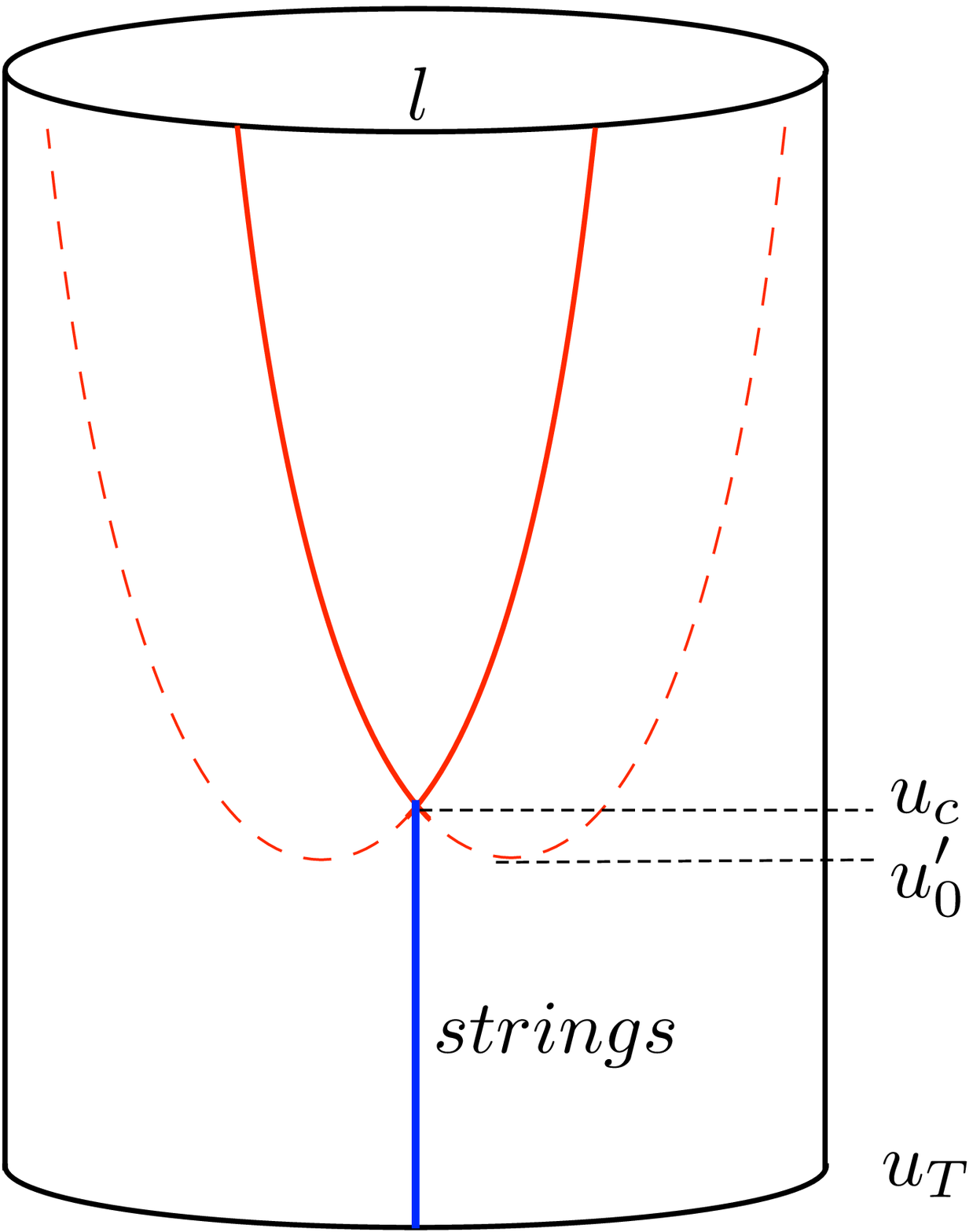,height=4cm}}
\caption{Possible 8-brane configurations with $d=0$ and $d\neq 0$ in the deconfined phase.}
\label{deconfined_configurations}
\end{figure}

The background describing the deconfined phase is given by exchanging the roles
of $x_4$ and $\tau$:
\be
\label{deconfined_background}
ds^2 = u^{3\over 2} \left(f(u)d\tau^2 + (d{\bf x})^2
+ dx_4^2\right)
+ u^{-{3\over 2}}\left({du^2\over f(u)}
+ u^2 d\Omega_4^2\right) \,,
\ee
with the same dilaton and RR 4-form as before, and where
\be
\label{thermal_factor}
f(u) = 1 - {u^3_T\over u^3} \;\; , \;\;
u_T = \left({4\pi\over 3}\right)^2 {R^2\over\beta_{\tau}^2} = \left({4\pi\over 3}\right)^2 t^2 \,,
\ee
where $t\equiv R/\beta_\tau = RT$ is the dimensionless temperature.
The 8-brane action is now given by
\be 
\label{D8_action_deconfined}
S_{D8} =  {\cal N}
\int du \, u^4 
\left[ f(u) (x_4^\prime(u))^2 
 + u^{-3} 
    \left( 1 - (\hat{a}_0^\prime(u))^2 \right)
\right]^{1\over 2} \,,
\ee
and the Legendre-transformed action is
\be
\label{transformed_D8_action_deconfined}
\tilde{S}_{D8} =  {\cal N} \int du\, u^4
\left[f(u)(x_4^\prime(u))^2 + u^{-3} \right]^{1\over 2} 
\left[ 1 + {(d(u))^2\over u^5} \right]^{1\over 2} \,,
\ee
where $d(u)$ is now given by
\be
\label{d_deconfined}
d(u) = {u \hat{a}_0^\prime(u)\over 
\left[ f(u) (x_4^\prime(u))^2 
 + u^{-3} \left(1 - (\hat{a}_0^\prime(u))^2 \right)
\right]^{1\over 2}} \,.
\ee
As in the confined phase, the equation of motion for $d(u)$ implies that it is a 
constant $d(u)=d$. 
On the other hand, for $x_4(u)$ there are two types of possible configurations
(figure \ref{deconfined_configurations}).
The first corresponds to separated parallel 8-branes and anti-8-branes with
\be
x_4'(u)=0 \,,
\ee
and the second to a connected configuration with
\be
\label{x4_solution_deconfined}
(x_4^\prime(u))^2 = {1\over u^3 f(u)}
\left[{f(u)(u^8+u^3d^2)\over f(u_0)(u_0^8+u_0^3d^2)} - 1 \right]^{-1} \,.
\ee
Actually, as we will soon see there are in fact two connected solutions, but
only one is (classically) stable.

\subsubsection{4-brane sources}

\begin{figure}
\centerline{\epsfig{file=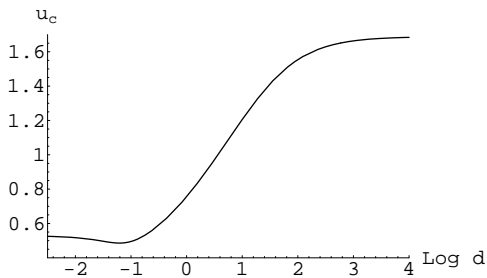,height=4cm}}
\caption{The position of the cusp in the 4-brane cusp configuration as a function of the 
electric displacement for $l=1$ in the deconfined phase.}
\label{deconfined_uc}
\end{figure}

The parallel configuration can have a uniform electric displacement $d$ without sources.
In the connected configuration, however, we need a source, as in the confined phase.
One possible source is again 4-branes inside the 8-branes.
The 4-brane action is now
\be
S_{D4} = {1\over 3} {\cal N} u_c \sqrt{f(u_c)}\, d \,,
\ee
and the force they exert is given by
\be
f_{D4} = {1\over 3} {\cal N} d \left(f(u_c) + {u_c f'(u_c)\over 2}\right) u_c^{3/4} 
= {1\over 3} {\cal N} d \frac{3-f(u_c)}{2} u_c^{3/4}\,,
\ee
where in the last equality we used the form of $f(u)$ in (\ref{thermal_factor}).
The 8-brane force is computed as before, but with the metric of the deconfined phase.
This gives
\be
f_{D8} = {\cal N} u_c^{13/4} \sqrt{f(u_c)} \left(1 + {d^2\over u_c^5}\right)^{1/2}\,,
\ee
and the same angle as before (\ref{cosine}).
The solution of the zero-force condition for a representative temperature is shown in 
figure \ref{deconfined_uc}.
The qualitative behavior is the same as in the confined phase:
initially the cusp comes down as $d$ increases, but eventually it goes
up and approaches a fixed value.

\begin{figure}
\centerline{\epsfig{file=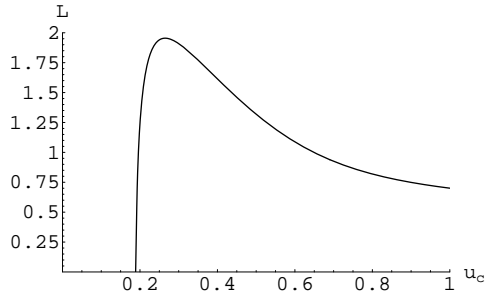,height=4cm}}
\caption{The asymptotic brane-anti-brane separation $l$ as a function of the cusp position $u_c$
for a fixed $d=0.5$ and $t=0.1$.}
\label{l_max}
\end{figure}

\begin{figure}
\centerline{\epsfig{file=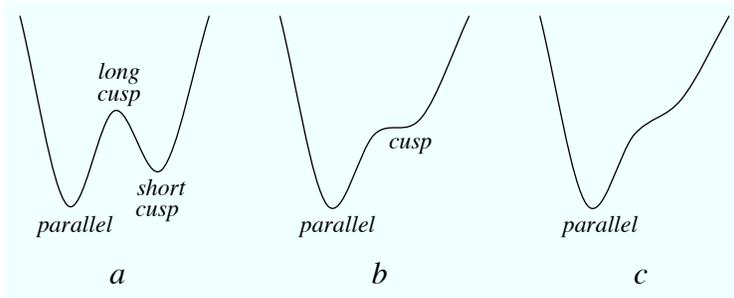,height=4cm}}
\caption{A schematic of the three possible solutions:
(a) $l<l_{max}$ (b) $l=l_{max}$ (c) $l>l_{max}$.}
\label{two_minima}
\end{figure}

In the deconfined phase there are actually two connected solutions in general.
This can be seen by looking at $l$ as a function of the cusp position $u_c$
at fixed $d$ and $t$ (figure \ref{l_max}). There are two values of $u_c$ for a given $l$ below
some $l_{max}$. At $l=l_{max}$ the two solutions coincide, and above $l_{max}$
there is no connected solution. This behavior is also true for $d=0$.
So in fact there are three solutions in all when $l<l_{max}$: the parallel configuration,
a ``short" cusp configuration (or ``short" U-configuration when $d=0$),
and a ``long" cusp configuration (or ``long" U-configuration when $d=0$).
When $l>l_{max}$ only the parallel configuration is a solution.
This picture is very reasonable from the following point of view. 
Imagine that we have two classically stable solutions in some theory
with a potential. The two solutions correspond to two local minima of the potential.
But this necessarily implies that there should be a third solution, corresponding
to the local maximum between the two minima (figure \ref{two_minima}a). 
This solution should be unstable.
Now imagine that one local minimum is lower than the other, and that the second
local minimum approaches the local maximum as we vary some parameter.
When they coincide we get a point of inflection (figure \ref{two_minima}b).
As we continue to vary the same parameter both solutions cease to exist,
leaving only the lower minimum (figure \ref{two_minima}c). 
This is precisely what happens for the 8-brane embedding.
The parallel and short cusp configurations are the stable solutions.
The long cusp configuration must therefore correspond to the unstable solution.
We leave it as a future excercise to exhibit the required negative mode.
Note that this picture necessarily implies that we don't have to worry
about the cusp solution disappearing, since in the region of parameter space
near this point the parallel solution always dominates.

\subsubsection{string sources}

The other possible sources of electric displacement in this phase are
strings which stretch from the 8-branes to the horizon at 
$u_T$.\footnote{Instead of ending at the horizon, 
the strings may also end on 4-branes which wrap the
$S^4$ and are located below the 8-branes. This is Witten's description of baryons 
\cite{Witten:1998xy}. Each 4-brane has $N_c$ strings attached to it.
However this configuration is not a solution of the equations of motion. 
There is a net force that pulls the 4-brane to larger $u$ \cite{Callan:1999zf}.
Eventually the 4-brane reaches the 8-brane and turns into an instanton.}
We can determine the precise relation between the density of strings $n_s$ and
the electric displacement $d$ by looking at the $B$-field dependence of the
supergravity, 8-brane, and string actions:
\be
S_{SUG}[B] &=& -{1\over 4\kappa_{10}^2}\int d^{10}x\,
\sqrt{-\mbox{det}g}\, e^{-2\Phi}\, |\partial B|^2 \\
S_{D8}[B] 
 &=& {\cal N} \int du \, u^4 \left[ f(u) (x_4^\prime(u))^2 
 + u^{-3} - u^{-3} \left(B_{0u} + \hat{a}_0'(u)\right)^2 \right]^{1\over 2}\\
S_{F1}[B] &=& -{n_s V_3\over 2\pi\alpha' R}\int d\tau du
\left(\sqrt{-\mbox{det}\, g_{MN}} - B_{0u}\right) \,.
\ee
Varying with respect to $B_{0u}$ and integrating over an 8-sphere surrounding
the endpoint of the strings in the 8-branes we find that 
\be
n_s = {2\pi\alpha' R^2{\cal N}\over \beta V_3}\, d \,.
\ee
Note that this is consistent with what we found for 4-branes
in (\ref{four_brane_density}),
since each 4-brane (away from the 8-branes) has $N_c$ strings attached.

\begin{figure}
\centerline{\epsfig{file=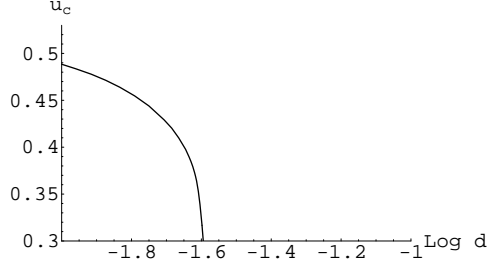,height=4cm}}
\caption{The position of the cusp in the stringy cusp configuration as a function of the 
electric displacement for $l=1$.}
\label{string_uc}
\end{figure}

Evaluating the string action for the deconfined background gives
\be
\label{NG_action}
S_{F1} = {\cal N} (u_c-u_T) d \,.
\ee
As in the 4-brane case, we have assumed a uniform distribution of strings in
$\mathbb{R}^3\times S^4$, so the point on the 8-brane where they end
will again be a cusp. 
The force downward applied by the strings is given by
\be
f_{F1} = {\delta S_{F1}\over \delta u_c} \left.{1\over\sqrt{g_{uu}}}\right|_{u_c}
= {\cal N} d\, u_c^{3/4} \sqrt{f(u_c)} 
\,.
\ee
The solution to the zero-force condition with the strings is shown in figure \ref{string_uc}.
The behavior is different from the 4-brane case.
The position of the cusp comes down monotonically with increasing $d$.

\begin{figure}
\centerline{\epsfig{file=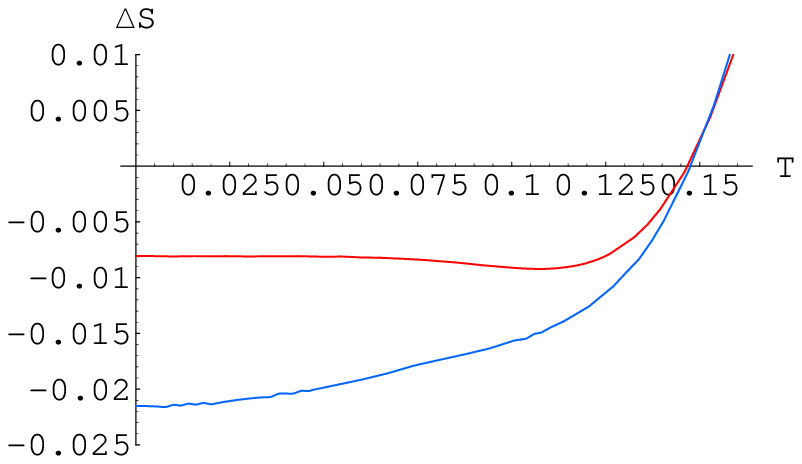,height=4cm}\hspace{1cm}
\epsfig{file=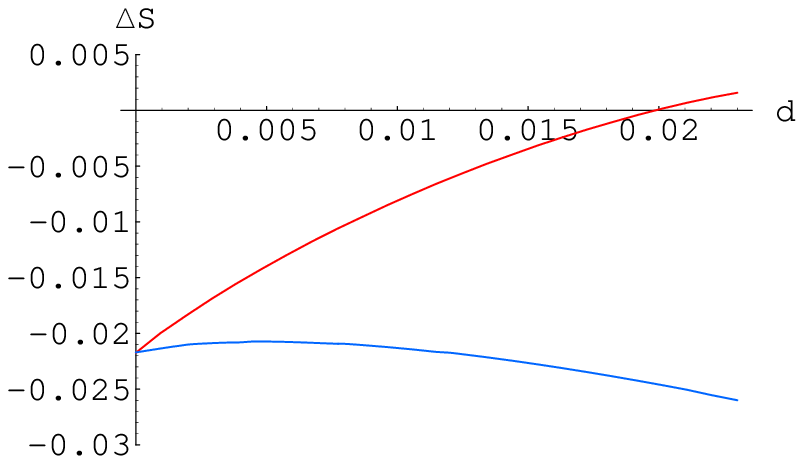,height=4cm}}
\caption{Comparing the actions (relative to the parallel configuration) 
of the string-sourced (red) and 4-brane-sourced (blue) cusp configurations.
The 4-brane case wins at all temperatures and all $d$.}
\label{string_vs_4brane}
\end{figure}

It turns out, however, that the stringy cusp configuration is always subdominant 
to the 4-brane cusp configuration.
We can see this by comparing their actions.
The total action will have a contribution from the 8-branes $\tilde{S}_{D8}$
given by (\ref{transformed_D8_action_deconfined}), where the integral is taken from $u_c$ to infinity,
and from either the 4-brane or string sources.
The integrals are divergent, but we can regularize them by subtracting the action of
the 8-branes in the parallel configuration in both cases.
The results are shown in figure \ref{string_vs_4brane}.
The action of the cusp configuration sourced by 4-branes is smaller than that
of the configuration sourced by strings at all temperatures and for all values of $d$.

In the next section we will show that in fact this configuration is unstable
to fluctuations in $d$. This is similar to the instability found in \cite{Kobayashi:2006sb}.
We will comment on a possible interpretation of this instability in the conclusions.


\section{Thermodynamics with finite chemical potential}

We now turn to the gauge theory implications of the configurations we found.
Our main goal is to understand the phase diagram of the gauge theory
at finite temperature $t$ and finite baryon chemical potential $\mu$.
Note that this model has an additional parameter not present in QCD 
corresponding to the asymptotic 8-brane-anti-8-brane separation $l$. 
We will generally fix $l=1$. We have also considered other 
values of $l$ (smaller and larger) and found no qualitative change in the results.

\subsection{baryon chemical potential}

The grand canonical potential is obtained by evaluating 
the 8-brane action (\ref{D8_action_confined}) or (\ref{D8_action_deconfined})
on the solution. For convenience we will normalize the potential by dividing out the normalization 
constant ${\cal N}$,
\be
\Omega(t,\mu) = {1\over{\cal N}}\, S_{D8}[t,x_4(u),\hat{a}_0(u)]_{solution} \,.
\ee
Note however that the potential, as well as all other thermodynamic
quantities associated to the matter, scale as $N_f N_c$.
The (dimensionless) baryon chemical potential $\mu$ is identified with 
the asymptotic value of the $U(1)$ gauge potential in the solution
\be
\label{chemical_potential}
\mu = \hat{a}_0(\infty) \,.
\ee
With our normalizations the baryon number density is given by\footnote{The true baryon
number density is given by (\ref{four_brane_density}).}
\be
n_b = \mbox{} - {\partial\Omega(t,\mu)\over\partial\mu} = d\,.
\ee
We will therefore use $d$ to denote also the density.

For computational purposes it is more convenient to 
express $\mu$ in terms of $d$ using the
canonical ensemble. The free energy is defined as
\be
F(t,d) = \Omega(t,\mu) + \mu d \,,
\ee
and the chemical potential is given by
\be
\mu = \left.{\partial F(t,d)\over \partial d}\right|_{t} \,.
\ee
The free energy is thus related to the Legendre-transformed 8-brane action on the solution.
In the cusp configuration the total free energy includes also the contribution of the source 4-branes or strings, evaluated at the position of the cusp:
\be
F(t,d) = {1\over{\cal N}} \left(\tilde{S}_{D8}[t,x_4(u),d(u)]_{solution} 
+  S_{source}(t,d,u_c)\right) \,.
\ee
The dependence on $d$ comes from three places:
the explicit dependence of $\tilde{S}_{D8}$ and $S_{source}$ on $d$,
the dependence of the solution for $x_4'$ on $d$, and the dependence on $d$ of 
$u_c$. Including all of these gives
\be
\mu & =& {1\over{\cal N}} \bigg\{
\int_{u_c}^\infty du\, \left.\left(
{\delta\tilde{S}_{D8}\over\delta d(u)}  +
{\delta\tilde{S}_{D8}\over\delta x_4'(u)}
{\partial x_4'\over\partial d} \right)\right|^{solution}_{t,l,u_c}\nonumber\\
&+& \left.{\partial u_c\over\partial d}\right|_{t,l}
\left. \left({\partial\tilde{S}_{D8}\over\partial u_c} +
{\partial S_{source}\over\partial u_c}\right)\right|^{solution}_{d,t,l}
+ \left.{\partial S_{source}\over \partial d}\right|_{t,l,u_c}
\bigg\} \,.
\ee
The second term vanishes since 
${\delta\tilde{S}_{D8}/\delta x_4'(u)}$ is constant by the equation of motion for $x_4$
and the integral of $\partial x_4'/\partial d$, at fixed $u_c$, gives
$\partial l/\partial d$ which vanishes since $l$ is fixed.
The third and fourth
terms cancel by the zero-force condition at the cusp (see Appendix), leaving
\be
\label{mu_general}
\mu = \int^\infty_{u_c} \hat{a}'_0(u)
+  {1\over{\cal N}} \left.{\partial S_{source}\over \partial d}\right|_{t,l,u_c} \,,
\ee
where $\hat{a}'_0(u)$ is related to $d$ by inverting the relation 
(\ref{d_confined}) or (\ref{d_deconfined}). 
The identification of the chemical potential with the value of the gauge potential
at infinity (\ref{chemical_potential}) therefore reflects a particular choice of gauge, in which
$\hat{a}_0(u_c)$ is identified with the mass of the source.
In the parallel configuration the source term vanishes, and the lower limit of
the integral is at the horizon $u=u_T$.
In this case the gauge choice (\ref{chemical_potential}) gives $\hat{a}_0(u_T)=0$,
which is consistent with the fact that the source becomes massless at the horizon.


\subsection{confined phase}

\begin{figure}
\centerline{\epsfig{file= 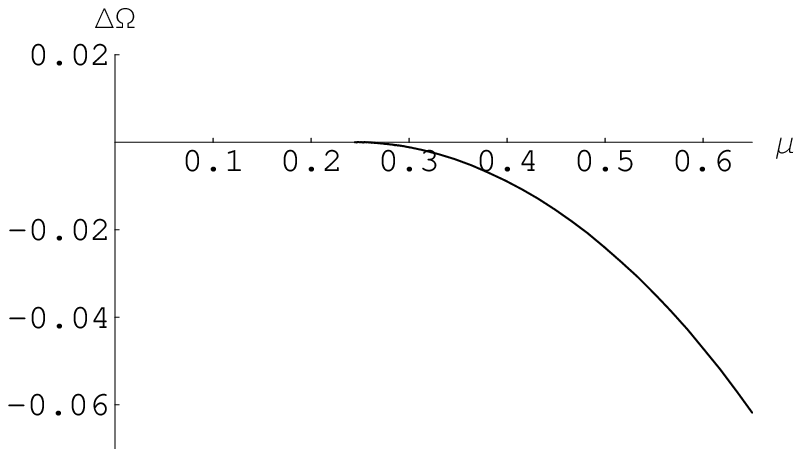,height=4cm}\hspace{1cm}
\epsfig{file=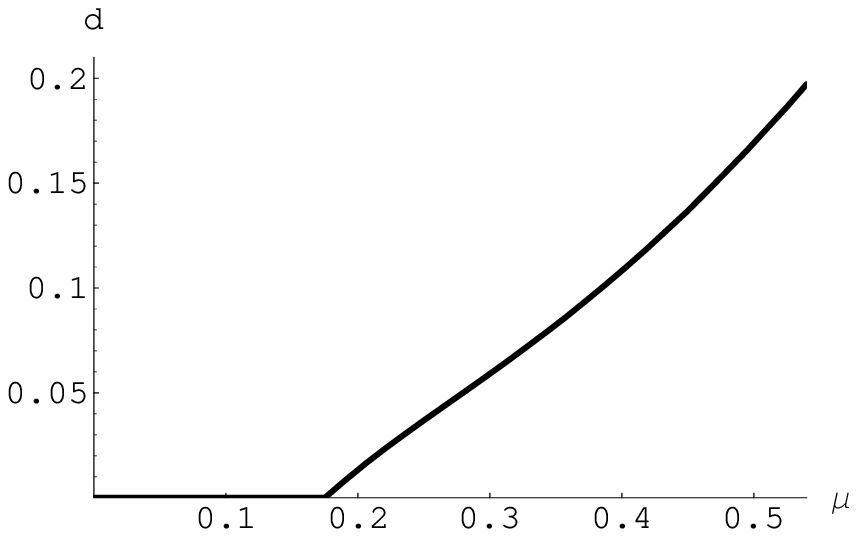,height=4cm}}
\caption{The difference in the grand canonical potential $\Omega$ between the nuclear 
and vacuum state, and the resulting baryon number density,
as a function of $\mu$ in the confined phase.} 
\label{nuclear_confined}
\end{figure}

In the confined phase only a connected 8-brane configuration is possible.
However for a given $\mu$, that is at fixed $\hat{a}_0(\infty)$, there are two 
connected solutions, 
a U-configuration with $d=0$ and a 4-brane sourced cusp configuration with 
$d\neq 0$.
The former corresponds to the QCD vacuum, and the latter to a phase of nuclear matter.
In the vacuum phase $\hat{a}_0$ is constant, and
the electric displacement $d$ vanishes. 
Therefore $\Omega$ does not depend on $\mu$ in this phase.
In the nuclear matter phase $\hat{a}_0(u)$ is sourced by 4-branes, and the chemical potential
is given by (\ref{mu_general}), which in the confined phase yields
\be
\label{mu_confined}
\mu = \int_{u_c}^{\infty} du \frac{d}{\sqrt{f(u) \left(u^5+d^2\right) - \left(\frac{u_0}{u}\right)^3 f(u_0) \left(u_0^5+d^2\right) }}+ \frac{1}{3} u_c \,.
\ee
Note that both $u_c$ and $u_0$ depend on $d$ and $l$.
There is no temperature dependence in the confined phase.
As we are working in the grand canonical ensemble, this represents an implicit
expression for $d(\mu)$. We see that there is a critical value for the chemical
potential $2u_c/3$ for which $d=0$. Below this value there is no cusp solution
and therefore no nuclear matter phase. This is precisely the onset chemical potential
$\mu_{onset}$. For $\mu > \mu_{onset}$ both the vacuum and the nuclear matter phases
exist, and we must compare their grand canonical potentials to determine which phase
is preferred. These quantities are actually divergent at $u\rightarrow\infty$,
but the difference is finite and is given by
\be
\Delta \Omega(\mu) &=& \Omega(\mu)_{nuc} - \Omega(\mu)_{vac} 
\nonumber\\
&=& \int_{u_c}^{\infty} \frac{u^{5/2}}{\sqrt{f(u) \left(1+\frac{d^2}{u^5}\right) 
- \frac{u_{0n}^8}{u^8} f(u_{0n})  \left(1+\frac{d^2}{u_{0n}^5}\right)}}  
-  \frac{u^{5/2}}{\sqrt{f(u) - \frac{u_{0v}^8}{u^8} f(u_{0v})}}  
\nonumber\\
& & -  \int_{u_{0v}}^{u_c} \frac{u^{5/2}}{\sqrt{f(u) - \frac{u_{0v}^8}{u^8} f(u_{0v})}}  \, ,
\ee
where $u_{0n}$ and $u_{0v}$ are the (different) values of $u_0$ in the nuclear and 
vacuum phases, respectively.
The results, for the representative value $u_{KK} = 0.5$ (and $l=1$), are shown in figure 
\ref{nuclear_confined}. 
Since $\Delta\Omega < 0 $ for all $\mu > \mu_{onset}$, the nuclear matter phase is preferred.  
Figure \ref{nuclear_confined} also shows the baryon number density as a function 
of the chemical potential for $u_{KK}=0.5$. (The value of $\mu_{onset}$
grows with $u_{KK}$).
Near $\mu=\mu_{onset}$ the density is continuous and behaves as 
$d\sim (\mu - \mu_{onset})^1$.
This is therefore a second-order phase transition with a critical exponent of 1.
This is a reasonable result given our approximation of ignoring the baryon interactions.




\subsection{deconfined phase}

In the deconfined phase the situation becomes more interesting since
there are more possible configurations at a given value of $\mu$.
In addition to the U-shaped and 4-brane-cusp configurations, there are the
parallel configuration, with vanishing or non-vanishing density, and the string-cusp configuration. 
The parallel configuration corresponds to a phase in which chiral symmetry 
is restored. At finite density this is the quark-gluon plasma (QGP).
The vacuum parallel configuration is irrelevant,
since it is clear from (\ref{D8_action_deconfined}) that a non-trivial $d$
is always preferred in the parallel configuration.
The string-cusp configuration features strings stretched between the 
8-branes and the horizon. We will refer to the corresponding phase
in the gauge theory as {\em quark matter}.
However, as we shall soon see
this configuration is actually unstable, at least for a uniform distribution
of baryon charge.
That leaves three phases to compare: the vacuum (U-configuration),
nuclear matter (4-brane-cusp) and the QGP (finite density parallel configuration).
We therefore expect in general a phase diagram in the $(t,\mu)$ plane
with three phase regions. 

\subsubsection{unstable quark matter}

\begin{figure}
\centerline{\epsfig{file= 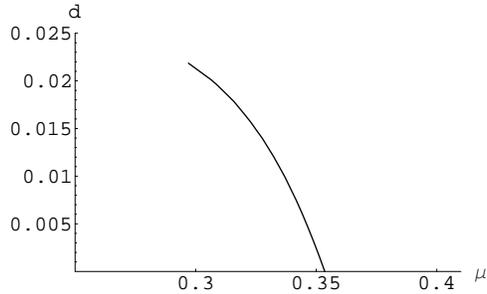,height=4cm}}
\caption{Baryon number density vs. chemical potential in the string-cusp configuration in the
deconfined phase.}
\label{string_density}
\end{figure}

Let us first show that the quark matter phase (string-cusp configuration) 
is thermodynamically unstable.
Evaluating the chemical potential (\ref{mu_general}) in the deconfined phase with string sources
(\ref{NG_action}) yields
\be
\mu = \int_{u_c}^{\infty} du 
\frac{\sqrt{f(u)}\, d}{\sqrt{f(u) \left(u^5+d^2\right) - \left(\frac{u_0}{u}\right)^3 f(u_0) \left(u_0^5+d^2\right) }}
+ (u_c - u_T) \,.
\ee
Figure \ref{string_density} shows a plot of $d$ vs. $\mu$ for this configuration.
It is apparent that
\be
{\partial d\over \partial \mu} < 0 \,,
\ee
(or equivalently $\partial\mu/\partial d < 0$ in the canonical ensemble)
and therefore that the string-cusp configuration is thermodynamically unstable to density
fluctuations. A similar instability was found in the D3-D7 model at finite density
\cite{Kobayashi:2006sb}. 

\subsubsection{phase diagram}

\begin{figure}
\centerline{\epsfig{file= 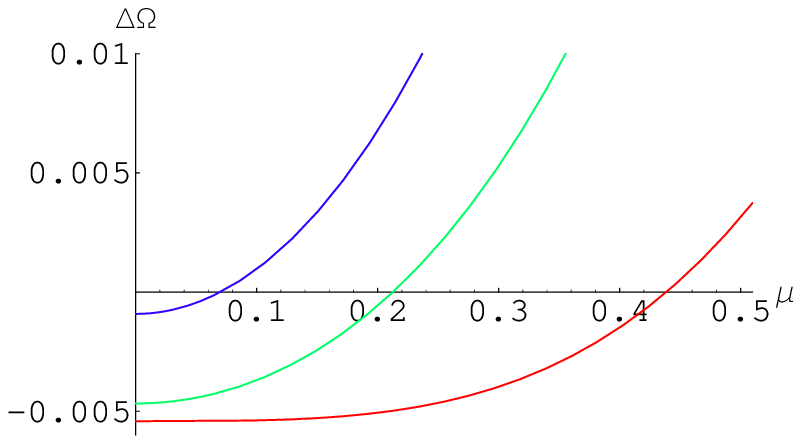,height=4cm}\hspace{1cm}
\epsfig{file= 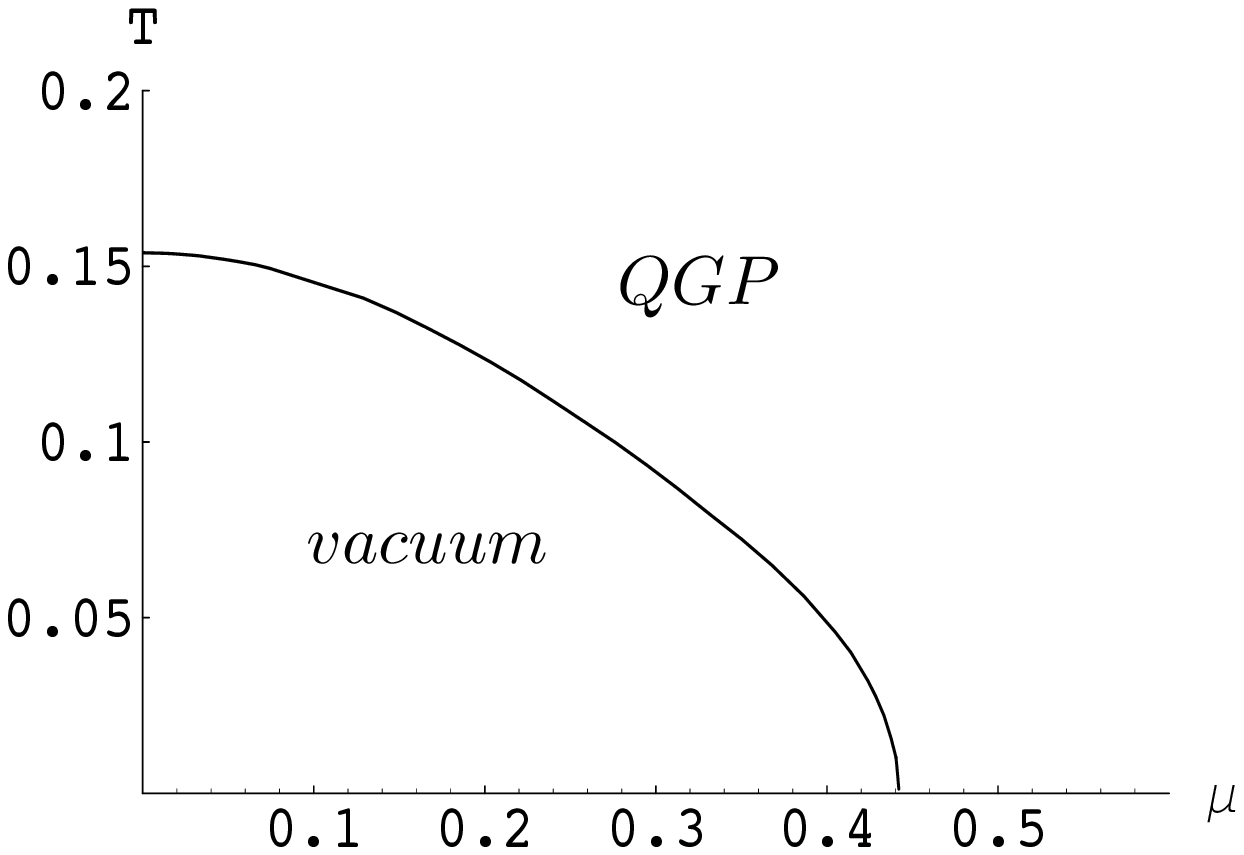,height=4cm}}
\caption{Grand canonical potential and phase diagram for the vacuum
vs. QGP phases.}
\label{U_vs_parallel}
\end{figure}

We begin by comparing the vacuum phase to the QGP phase.
The vacuum phase is described by the U-configuration, and the quark-gluon plasma
(QGP) phase is described by the parallel configuration.
The grand canonical potential of the vacuum can be read off from 
(\ref{D8_action_deconfined}) and (\ref{x4_solution_deconfined}) with 
$d=0$:
\be
\Omega_{vac}(\mu) = \int_{u_0}^\infty du\,
{u^{5/2}\,\sqrt{f(u)}\over \sqrt{f(u) - {u_0^8\over u^8} f(u_0)}} \,.
\ee
The potential of the QGP is given by (\ref{D8_action_deconfined})
and (\ref{d_deconfined}) with $x_4'(u)=0$:
\be
\Omega_{qgp}(\mu) = \int_{u_T}^\infty du\,
{u^5\over\sqrt{u^5 + d^2}} \,.
\ee
The density $d$ is a function of $\mu$ which is obtained from 
(\ref{mu_general}) without sources, and (\ref{d_deconfined}).
This gives
\be
\mu = \int_{u_T}^\infty du\,
{d\over\sqrt{u^5 + d^2}} \,,
\ee
which can be inverted numerically to get $d(\mu)$.
Both potentials are divergent at $u\rightarrow\infty$ but the difference is finite:
\be
\Delta\Omega_1 &=& \Omega_{qgp} - \Omega_{vac} \,.
\ee
Figure \ref{U_vs_parallel} shows $\Delta\Omega_1(\mu)$ for a few 
representative temperatures (and $l=1$).
The transition between the two phases occurs when $\Delta\Omega_1 = 0$.
We find a line of transitions between these two phases in the $(t,\mu)$ plane
shown in figure \ref{U_vs_parallel}.
This result was obtained previously in \cite{Horigome:2006xu}.

\begin{figure}
\centerline{\epsfig{file= 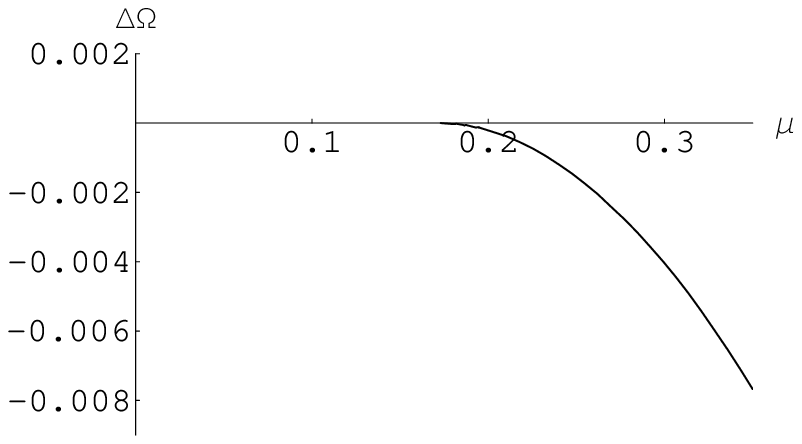,height=4cm}\hspace{1cm}
\epsfig{file= 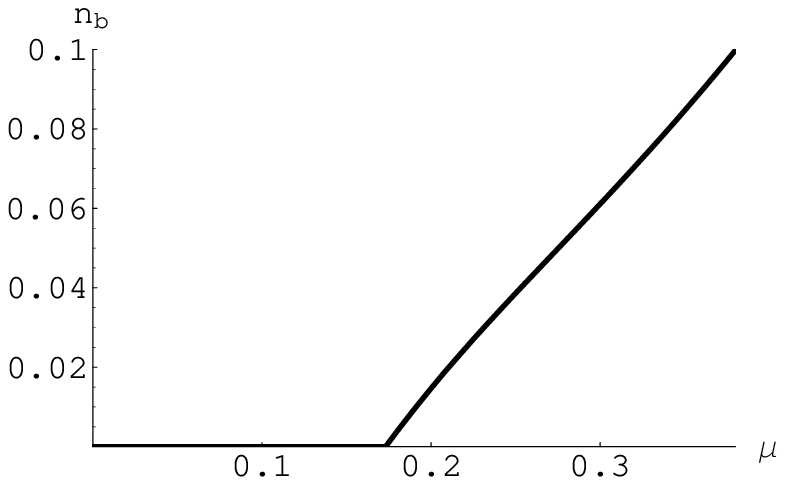,height=4cm}}
\caption{Grand canonical potential and baryon number density in the nuclear matter phase
relative to the vacuum phase.}
\label{U_vs_cusp_1}
\end{figure}
\begin{figure}
\centerline{\epsfig{file= 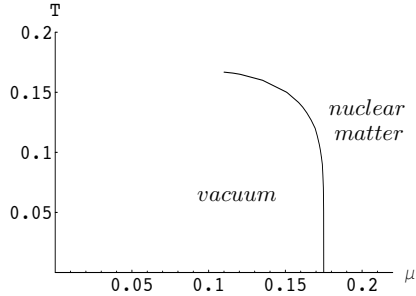,height=4cm}}
\caption{Phase diagram for vacuum and nuclear matter phases.}
\label{U_vs_cusp_2}
\end{figure}

We now turn to the comparison of the vacuum phase with the nuclear matter phase.
We did this already in the confined phase, and found a second-order phase transition
at some $\mu=\mu_{onset}$.
Since the Sakai-Sugimoto model exhibits chiral-symmetry breaking also in the deconfined
phase, it is reasonable to expect that nuclear matter should form also in this case.
The potential of the nuclear phase (with 4-branes) is given by
\be
\Omega_{nuc}(\mu) = \int_{u_c}^\infty du\,
{u^{5/2}\, \sqrt{f(u)}\over
\sqrt{f(u)\left(1+{d^2\over u^5}\right) 
- {u_0^8\over u^8}f(u_0)\left(1 + {d^2\over u_0^5}\right)}} \,,
\ee
where $d$ is again given implicitly in terms of $\mu$ using (\ref{mu_general}) with 4-brane sources:
\be
\mu =  \int_{u_c}^{\infty} du 
\frac{\sqrt{f(u)}\, d}{\sqrt{f(u) \left(u^5+d^2\right) - \left(\frac{u_0}{u}\right)^3 f(u_0) 
\left(u_0^5+d^2\right) }}
+ {1\over 3}u_c\sqrt{f(u_c)} \,.
\ee
We are now interested in the difference between the potentials of the nuclear phase and the
vacuum phase,
\be
\Delta\Omega_2 = \Omega_{nuc} - \Omega_{vac} \,.
\ee
Figure \ref{U_vs_cusp_1} shows $\Delta\Omega_2(\mu)$ for a representative temperature.
The behavior is qualitatively the same at all temperatures: $\Delta\Omega_2$
is negative for all $\mu$ for which the nuclear phase exists.
Figure \ref{U_vs_cusp_1} also shows the density as a function of $\mu$ at the same temperature.
As in the confined phase, the critical exponent is 1 to within our numerical accuracy, 
so the transition is second
order. By varying the temperature we obtain the phase diagram in figure \ref{U_vs_cusp_2}.
The behavior agrees qualitatively with what is expected in QCD: $\mu_{onset}$
decreases slightly as the temperature increases. 

\begin{figure}
\centerline{\epsfig{file= 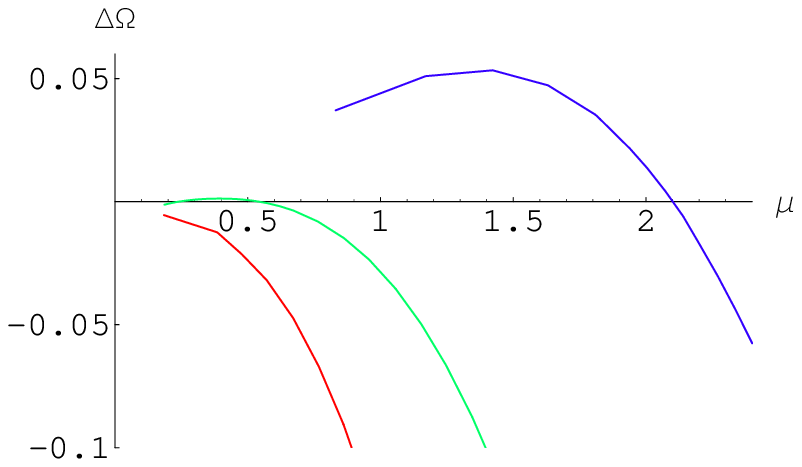,height=4cm}\hspace{1cm}
\epsfig{file=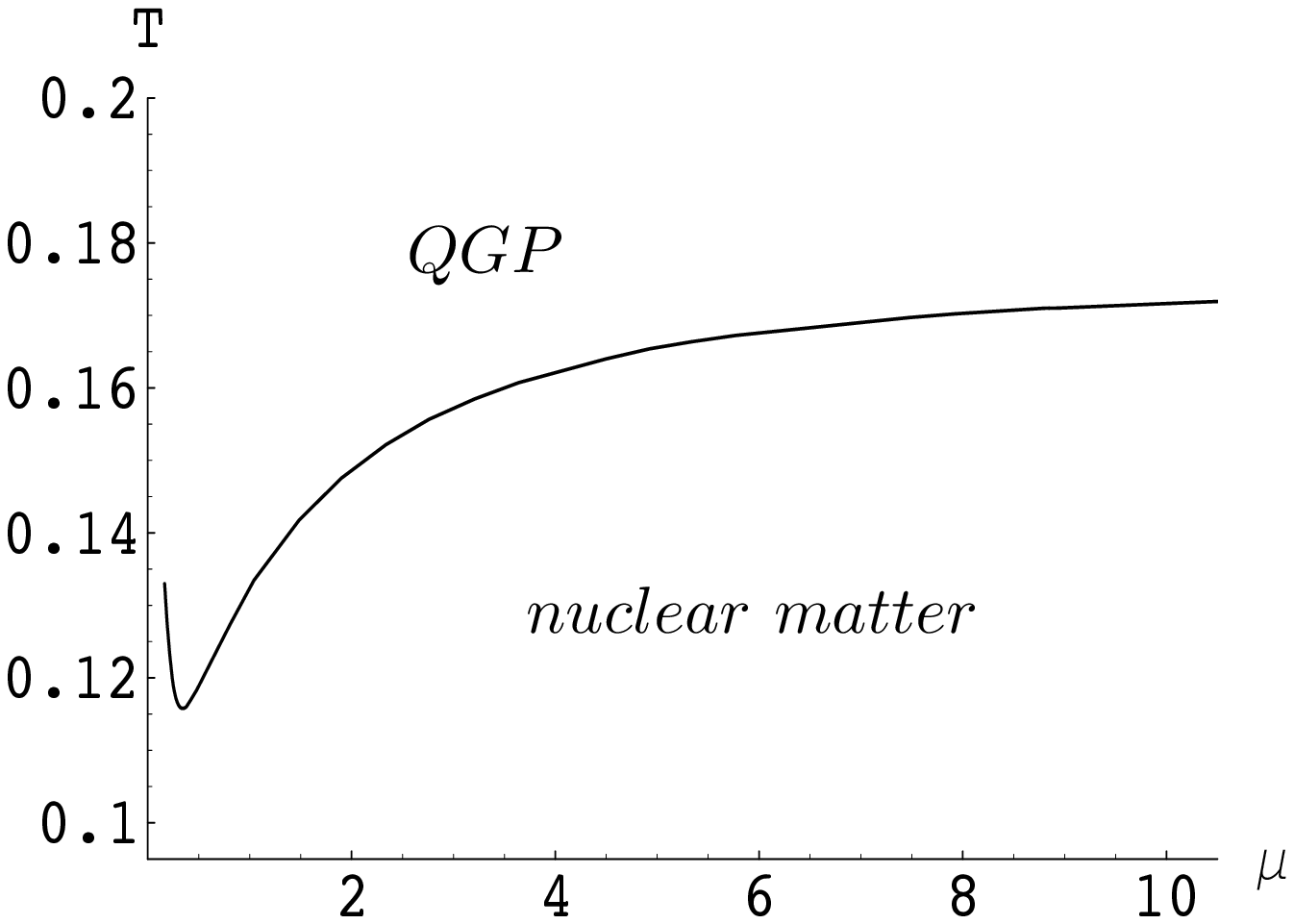,height=4cm}}
\caption{Grand canonical potential (for $t=0.1,0.12,0.15$) and phase diagram
for the nuclear vs. QGP phases.}
\label{cusp_vs_parallel}
\end{figure}

The final part of the phase diagram comes from comparing the nuclear and QGP phases:
\be
\Delta\Omega_3 = \Omega_{nuc} - \Omega_{qgp} \,.
\ee
Here we find an interesting temperature dependence.
Figure \ref{cusp_vs_parallel} shows $\Delta\Omega_3(\mu)$
for three representative temperatures.
At low temperature the nuclear matter phase wins for all $\mu$. Then
there is a temperature range for which the system undergoes two transitions as $\mu$
is increased,
first from nuclear matter to QGP, and then back to nuclear matter.
The resulting phase diagram is shown in figure \ref{cusp_vs_parallel}.
The physical source of the dip in the phase diagram is the dip that occurs
in the position of the cusp $u_c$ as a function of the density $d$ 
(figure \ref{deconfined_uc}).
There is a similar dip in the phase diagram of QCD
(see for example \cite{Rajagopal}).
At high temperature (not shown) the QGP phase is preferred for all $\mu$.

Finally, combining the three separate phase diagrams gives the complete phase diagram
shown in figure 1. At low temperature and chemical potential the vacuum phase dominates,
at low temperature and high chemical potential the nuclear phase dominates,
and at high temperature chiral symmetry is restored and the quark-gluon plasma
phase dominates.


\subsection{entropy and equation of state}

Phases of thermodynamic systems are also characterized by the 
their equation of state and entropy.
Let us briefly discuss these for the different phases we have encountered.

\begin{figure}
\centerline{\epsfig{file=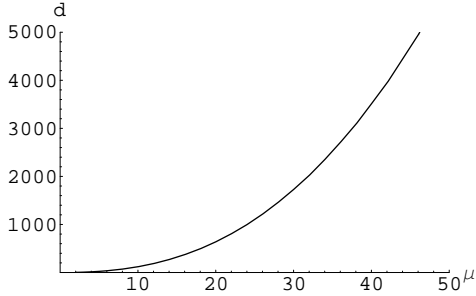,height=4cm}}
\caption{Density vs. chemical potential for large densities.}
\label{large_density}
\end{figure}

The pressure as a function of the density $p(d,t)$ is essentially given by 
$-\Omega(\mu(d),t)$. 
We find that at low temperature the behavior in the confined and
deconfined phases is very similar.\footnote{The behavior at high temperature, {\em i.e.}
in the QGP phase, was essentially worked out in \cite{PS}.}
At small densities $d\sim (\mu - \mu_{onset})$ and therefore\footnote{For a free fermi gas  
$p(d) \sim d^{5/3}$.}
\begin{equation}
p(d) \sim (\mu-\mu_{onset})^2 \sim d^2 \,.
\end{equation}
At large densities $d\sim \mu^{5/2}$ (figure \ref{large_density}) and thus
\begin{equation}
p(d)  \sim \mu^{7/2} \sim d^{7/5} \,.
\end{equation}
It is interesting that although we have not specified that the baryons are 
fermions (indeed there seem to be both fermionic and bosonic components),
the results for $\mu(d)$ mimic a behavior expected for fermions. This is due to the 
response of the DBI action to the electric field.

\begin{figure}
\centerline{\epsfig{file=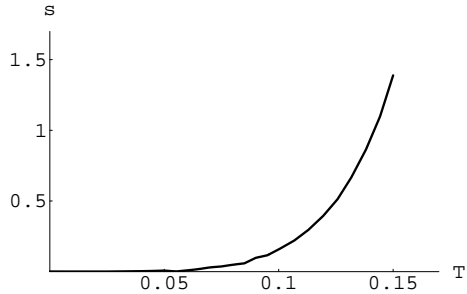,height=4cm}}
\caption{Entropy vs. temperature in the deconfined phase.}
\end{figure}

The entropy as a function of the temperature $s(t)$ is computed
from the free-energy $F(t,d)$.
The interesting case is the deconfined phase,
since there is no temperature dependence in the confined phase.
At low temperature, where chiral symmetry is broken, we find
(for both small and large densities)
\begin{equation}
s(t) \sim t^{5} \,,
\end{equation}
and at the high temperature, where chiral symmetry is restored, we find
\begin{equation}
s(t) \sim t^{6}.
\end{equation}


\section{Conclusions}

In this paper we have analyzed the different phases of the Sakai-Sugimoto model
at finite temperature and baryon chemical potential and determined 
the phase diagram.
In many respects our phase diagram is similar to that of QCD.
In both cases chiral symmetry is broken at low temperature and restored at high 
temperature, at all values of the chemical potential.
The dip in the phase diagram suggests that the chiral condensate initially 
decreases with $\mu$ and then increases.
This is similar to the behavior in QCD.
It would be interesting to study this directly using the holographic description
of the chiral condensate in terms of the tachyon \cite{BSS}.

The finite density phase is described by a gas of "baryonic matter" 
4-branes wrapped on the $S^4$ inside the 8-branes. 
At low temperatures this phase always dominates over the quark-gluon plasma phase.
The other possible description of baryon matter in terms of strings
("quark matter") turns out to be subdominant and unstable.
If we ignore these facts and use strings instead of 4-branes to describe
baryonic matter the phase diagram would change and chiral symmetry
would be restored at high density.

We also find a phase transition between the vacuum and nuclear matter phases.
In QCD this is a first-order transition, but in our case it is second-order.
We believe that that the difference is a result
of neglecting the interactions between the 4-branes. 

Another difference is that QCD at high density is expected to be in a CFL phase,
in which both the chiral symmetry and the gauge symmetry are broken.
However, at large $N_c$ QCD is expected to be dominated by
a non-uniform chiral symmetry breaking phase with unbroken
gauge symmetry.
We did not explore this possibility in this paper.
However the result that the ``quark matter" phase,
with strings stretched to the horizon, was unstable to density fluctuations,
suggests that there may exist a stable non-unifrom phase.
It would be interesting to see if it is similar to the chiral density wave 
in large $N_c$ QCD.


\appendix
\section{Zero-force condition from the action}

The force balance condition for the cusp configurations can alternately be obtained
directly by varying the total action with respect to the cusp position $u_c$.
The total action is given by
\begin{equation}
\tilde{S}=\tilde{S}_{D8}+
S_{source}(u_c)
=\int_{u_{c}}^{\infty} \tilde{L} (x_{4}'(u),d,t) du +
S_{source}(u_c,d,t) \,.
\end{equation}
We want to vary the action with respect to $u_c$, while keeping the physical varaibles $l,t$ and $d$
fixed. 
To do this we need to vary $x_{4}'$ (and therefore $u_0$) accordingly. 
This gives
\begin{equation}
\left.{\partial\tilde{S}\over\partial u_{c}}\right|_{d,t,l} = 
-\tilde{L}(u_c)+\int_{u_c}^{\infty} du\,
\left.{\delta \tilde{S}_{D8}\over\delta x_{4}'}
{\partial x_{4}'\over \partial u_c}\right|_{d,t,l} +
\left.{\partial S_{source}\over\partial u_c}\right|_{d,t,l} \,.
\end{equation}
However since $l$ is given by
\begin{equation}
l= 2 \int_{u_c}^{\infty} du\,  x_{4}'(u)
\end{equation}
we get
\begin{equation}
-x_{4}'(u_{c})+\int_{u_c}^{\infty} du\, \left.\frac{\partial x_{4}^{'}}{\partial u_{c}}\right|_{d,t,l}=0 \,.
\end{equation}
Furthermore, the equation of motion sets $\delta\tilde{S}_{D8}/\delta x_4'$ to a constant
independent of $u$.
Requiring the total action to be stationary with respect to the varaition of $u_c$ one gets
\begin{equation}
\tilde{L}(u_c)-x_{4}'(u_c) \frac{\delta \tilde{S}_{D8}}{\delta x_{4}'}
=\frac{\partial S_{source}}{\partial u_{c}} \,.
\end{equation}
Substituting in the expressions for $\tilde{L}$ and $S_{source}$ 
then reproduces the force balance conditions in the various cases
(confined, deconfined, 4-branes, strings).

\section*{Acknowledgments}
We would like to thank Ofer Aharony, David J. Bergman, Sumit Das, Ben Freivogel, 
Yariv Kafri, Keh-Fei Liu, David Mateos, Rob Myers, Al Shapere, Cobi Sonnenschein 
and Shigeki Sugimoto for helpful conversations.
This work was supported in part by the
Israel Science Foundation under grant no.~568/05.


\begin{thebibliography}{99}

\bibitem{Schafer:2005ff}
  T.~Schafer,
  arXiv:hep-ph/0509068.
\bibitem{Alford}
  M.~G.~Alford,
  PoS {\bf LAT2006}, 001 (2006)
  [arXiv:hep-lat/0610046].
\bibitem{Rajagopal}
  K.~Rajagopal and F.~Wilczek,
  arXiv:hep-ph/0011333.
  
\bibitem{Alford:1998mk}
  M.~G.~Alford, K.~Rajagopal and F.~Wilczek,
  Nucl.\ Phys.\  B {\bf 537}, 443 (1999)
  [arXiv:hep-ph/9804403].
  
  \bibitem{dgr}
  D.~V.~Deryagin, D.~Y.~Grigoriev and V.~A.~Rubakov,
  Int.\ J.\ Mod.\ Phys.\  A {\bf 7} (1992) 659.

\bibitem{ss}
  E.~Shuster and D.~T.~Son,
  Nucl.\ Phys.\  B {\bf 573}, 434 (2000)
  [arXiv:hep-ph/9905448].
  
  
\bibitem{Lombardo:2006yc}
  M.~P.~Lombardo,
  PoS C {\bf POD2006}, 003 (2006)
  [arXiv:hep-lat/0612017].
  
  

\bibitem{Aharony:1999ti}
  O.~Aharony, S.~S.~Gubser, J.~M.~Maldacena, H.~Ooguri and Y.~Oz,
  Phys.\ Rept.\  {\bf 323}, 183 (2000)
  [arXiv:hep-th/9905111].
\bibitem{Flavors}
A.~Karch and E.~Katz,
  JHEP {\bf 0206}, 043 (2002)
  [arXiv:hep-th/0205236];
T.~Sakai and J.~Sonnenschein,
  JHEP {\bf 0309}, 047 (2003)
 [arXiv:hep-th/0305049];
J.~Babington, J.~Erdmenger, N.~J.~Evans, Z.~Guralnik and I.~Kirsch,
  Phys.\ Rev.\  D {\bf 69}, 066007 (2004)
  [arXiv:hep-th/0306018];
X.~J.~Wang and S.~Hu,
  JHEP {\bf 0309}, 017 (2003)
  [arXiv:hep-th/0307218];
P.~Ouyang,
  Nucl.\ Phys.\  B {\bf 699}, 207 (2004)
  [arXiv:hep-th/0311084];
  C.~Nunez, A.~Paredes and A.~V.~Ramallo,
  JHEP {\bf 0312}, 024 (2003)
  [arXiv:hep-th/0311201];
 M.~Kruczenski, D.~Mateos, R.~C.~Myers and D.~J.~Winters,
  JHEP {\bf 0405}, 041 (2004)
  [arXiv:hep-th/0311270];
  N.~J.~Evans and J.~P.~Shock,
  Phys.\ Rev.\  D {\bf 70}, 046002 (2004)
  [arXiv:hep-th/0403279];
R.~Casero, A.~Paredes and J.~Sonnenschein,
  JHEP {\bf 0601}, 127 (2006)
  [arXiv:hep-th/0510110].
  

\bibitem{Sakai:2004cn}
  T.~Sakai and S.~Sugimoto,
  Prog.\ Theor.\ Phys.\  {\bf 113}, 843 (2005)
  [arXiv:hep-th/0412141].
\bibitem{Witten:1998zw}
  E.~Witten,
  Adv.\ Theor.\ Math.\ Phys.\  {\bf 2}, 505 (1998)
  [arXiv:hep-th/9803131].

\bibitem{SS_model_general}
T.~Sakai and S.~Sugimoto,
  Prog.\ Theor.\ Phys.\  {\bf 114}, 1083 (2006)
  [arXiv:hep-th/0507073];
K.~Peeters, J.~Sonnenschein and M.~Zamaklar,
  JHEP {\bf 0602}, 009 (2006)
  [arXiv:hep-th/0511044];
  P.~Benincasa and A.~Buchel,
  Phys.\ Lett.\  B {\bf 640}, 108 (2006)
  [arXiv:hep-th/0605076];
  K.~Peeters, J.~Sonnenschein and M.~Zamaklar,
  Phys.\ Rev.\  D {\bf 74}, 106008 (2006)
  [arXiv:hep-th/0606195];
  Y.~h.~Gao, W.~s.~Xu and D.~f.~Zeng,
  arXiv:hep-th/0611217.
   K.~Nawa, H.~Suganuma and T.~Kojo,
  Phys.\ Rev.\  D {\bf 75}, 086003 (2007)
  [arXiv:hep-th/0612187];
 O.~Bergman and G.~Lifschytz,
  JHEP {\bf 0704}, 043 (2007)
  [arXiv:hep-th/0612289];
  K.~Nawa, H.~Suganuma and T.~Kojo,
  arXiv:hep-th/0701007;
 D.~K.~Hong, M.~Rho, H.~U.~Yee and P.~Yi,
  arXiv:hep-th/0701276;
   K.~Hashimoto, T.~Hirayama and A.~Miwa,
  JHEP {\bf 0706}, 020 (2007)
  [arXiv:hep-th/0703024];
 D.~K.~Hong, M.~Rho, H.~U.~Yee and P.~Yi,
  arXiv:0705.2632 [hep-th].

\bibitem{Hata:2007mb}
  H.~Hata, T.~Sakai, S.~Sugimoto and S.~Yamato,
  arXiv:hep-th/0701280;

\bibitem{Aharony:2006da}
  O.~Aharony, J.~Sonnenschein and S.~Yankielowicz,
  arXiv:hep-th/0604161.





\bibitem{Kim:2006gp}
  K.~Y.~Kim, S.~J.~Sin and I.~Zahed,
  arXiv:hep-th/0608046;
\bibitem{Horigome:2006xu}
  N.~Horigome and Y.~Tanii,
  JHEP {\bf 0701}, 072 (2007)
  [arXiv:hep-th/0608198].
\bibitem{Sin:2007ze}
  S.~J.~Sin,
  arXiv:0707.2719 [hep-th].
\bibitem{Yamada:2007ys}
  D.~Yamada,
  arXiv:0707.0101 [hep-th].
  
  \bibitem{Kobayashi:2006sb}
  S.~Kobayashi, D.~Mateos, S.~Matsuura, R.~C.~Myers and R.~M.~Thomson,
  JHEP {\bf 0702}, 016 (2007)
  [arXiv:hep-th/0611099].
  
  
  
\bibitem{Domokos:2007kt}
  S.~K.~Domokos and J.~A.~Harvey,
  arXiv:0704.1604 [hep-ph].
  
\bibitem{Kim:2007em}
  Y.~Kim, B.~H.~Lee, S.~Nam, C.~Park and S.~J.~Sin,
  arXiv:0706.2525 [hep-ph].

\bibitem{Kim:2007xi}
  Y.~Kim, C.~H.~Lee and H.~U.~Yee,
  arXiv:0707.2637 [hep-ph].
  
  


\bibitem{NJL}
  E.~Antonyan, J.~A.~Harvey, S.~Jensen and D.~Kutasov,
  arXiv:hep-th/0604017;
  A.~Parnachev and D.~A.~Sahakyan,
  Phys.\ Rev.\ Lett.\  {\bf 97}, 111601 (2006)
  [arXiv:hep-th/0604173];
\bibitem{PS}  
 A.~Parnachev and D.~A.~Sahakyan,
  Nucl.\ Phys.\  B {\bf 768}, 177 (2007)
  [arXiv:hep-th/0610247].
  
  
  
\bibitem{Witten:1998xy}
  E.~Witten,
  JHEP {\bf 9807}, 006 (1998)
  [arXiv:hep-th/9805112].
  
  
\bibitem{Callan:1999zf}
  C.~G.~.~Callan, A.~Guijosa, K.~G.~Savvidy and O.~Tafjord,
  Nucl.\ Phys.\  B {\bf 555}, 183 (1999)
  [arXiv:hep-th/9902197].
  
\bibitem{BSS}
O.~Bergman, J.~Sonnenschein, S.~Seki, to appear.


  
\end{thebibliography}
\end{document}